\documentclass[pre,twocolumn,showpacs]{revtex4-1}

\usepackage{amssymb,amsmath}

\usepackage{graphicx}

\newcommand\beq{\begin{equation}}
\newcommand\eeq{\end{equation}}
\newcommand\beqa{\begin{eqnarray}}
\newcommand\eeqa{\end{eqnarray}}
\newcommand{\dd}{\text{d}}
\newcommand{\nn}{\nonumber\\}

\newcommand{\al}{\alpha}

\usepackage{color}
\newcommand{\rojo}[1]{{ #1}}

\begin{document}

\title{Transport coefficients for driven granular mixtures at low-density}
\author{Nagi Khalil}
\email{nagi@us.es}
\affiliation{Departamento de
F\'{\i}sica, Universidad de Extremadura, E-06071 Badajoz, Spain}
\author{Vicente Garz\'o}
\email{vicenteg@unex.es} \homepage{http://www.unex.es/eweb/fisteor/vicente/} \affiliation{Departamento de
F\'{\i}sica, Universidad de Extremadura, E-06071 Badajoz, Spain}

\begin{abstract}
The transport coefficients of a granular binary mixture driven by a stochastic bath with friction are determined from the inelastic Boltzmann kinetic equation. A normal solution is obtained via the Chapman-Enskog method for states near homogeneous steady states. The mass, momentum, and heat fluxes are determined to first order in the spatial gradients of the hydrodynamic fields, and the associated transport coefficients are identified. They are given in terms of the solutions of a set of coupled linear integral equations. As in the monocomponent case, since the collisional cooling cannot be compensated locally for by the heat produced by the external driving, the reference distributions (zeroth-order approximations) $f_i^{(0)}$ ($i=1,2$) for each species depend on time through their dependence on the pressure and the temperature. Explicit forms for the diffusion transport coefficients and the shear viscosity coefficient are obtained by assuming the steady state conditions and by considering the leading terms in a Sonine polynomial expansion. A comparison with previous results obtained for granular Brownian motion and by using a (local) stochastic thermostat is also carried out. The present work extends previous theoretical results derived for monocomponent dense gases [V. Garz\'o, M. G. Chamorro, and F. Vega Reyes, Phys. Rev. E \textbf{87}, 032201 (2013)] to granular mixtures at low density.
\end{abstract}


\date{\today}
\maketitle

\section{Introduction}
\label{sec1}

The use of kinetic theory to describe granular matter under rapid flow conditions (i.e., when material is externally excited) has been an active area of research in the past several decades \cite{G03,BP04}. On the other hand, although  in many conditions the motion of grains exhibits a great similarity to the random motion of atoms or molecules of an ordinary gas, the fact that collisions between grains are inelastic gives rise to subtle modifications of the conventional hydrodynamic equations. In particular, since the energy is decreasing with time, one has to feed energy into the system to keep it under rapid flow conditions. When the injected energy compensates for the energy lost by collisions, a non-equilibrium \emph{steady} state is achieved. In this sense, granular matter can be seen as a good example of a system which is inherently in a non-equilibrium state.

In real experiments, the energy input can be done either by driving through the boundaries \cite{boundaries} or alternatively by bulk driving, as in air-fluidized beds \cite{AD06,SGS05}. However, these ways of supplying energy produces in many cases strong spatial gradients in the bulk domain. The same effect can be reached by heating the system homogenously by the action of an external driving force. This is the usual way to drive a granular gas in computer simulations \cite{puglisi,ernst}. Borrowing a terminology used in non-equilibrium molecular dynamics simulations of ordinary fluids \cite{EM90}, this type of external forces are called ``thermostats''. Although thermostats have been widely used in the past to analyze granular flows, their influence on the properties of the system is still an unsolved problem, even in the case of ordinary fluids \cite{DSBR86,GSB90,GS03}.

The transport coefficients of a driven granular monodisperse fluid have been recently determined \cite{GCV13}. In this work, the fluid is driven by the action of a thermostat that is composed by two terms: (i) a drag force proportional to the velocity of the particle and (ii) a stochastic force with the form of a Gaussian white noise where the particles are randomly kicked between collisions \cite{WM96}. While the viscous drag force could model the friction of grains with a surrounding fluid (interstitial gas phase), the stochastic force could model the energy transfer from the interstitial fluid molecules to the granular particles. At a kinetic level, the results derived in Ref.\ \cite{GCV13} were obtained by solving the (inelastic) Enskog equation by means of the Chapman-Enskog (CE) method to first order in the spatial gradients (Navier-Stokes hydrodynamic order). Thus, these results go beyond the dilute regime and apply in principle to moderate densities where the collisional contributions to the fluxes cannot be neglected. The kind of thermostat used in Ref.\ \cite{GCV13} has been widely used in previous works by other authors to perform computer simulations\cite{puglisi}. Moreover, it must be remarked that the model (stochastic bath with friction) has been also shown to be relevant in more practical applications since some recent experimental results for structure factors \cite{GSVP11,PGGSV12} can be fairly well reproduced by the present model.

Nevertheless, real granular systems are usually present in nature as \emph{multicomponent} systems, namely, they are constituted by particles of different mechanical properties. Therefore, a very interesting problem is to extend the results derived  for a monocomponent granular gas in Ref.\ \cite{GCV13} to the case of granular mixtures. On the other hand, the analysis of transport phenomena in fluid mixtures is much more complicated than for monocomponent gases. Not only is the number of transport coefficients higher but also these coefficients depend on more parameters such as the volume fractions, concentrations, masses, sizes, and/or coefficients of restitution. Thus, in order to gain some insight into the general problem, one considers first more simple systems such as the case of granular binary mixtures at low-density.

The goal of this paper is to evaluate the transport coefficients of a dilute granular binary mixture driven by a stochastic bath with friction. As in the undriven case \cite{GD02}, the transport coefficients are obtained by solving the set of coupled nonlinear Boltzmann equations by means of the CE method \cite{CC70} conveniently adapted to account for the inelastic character of collisions. However, while in the undriven case the zeroth-order approximations $f_i^{(0)} (i=1,2)$ of each species are chosen to be the \emph{local} version of the so-called homogeneous cooling state (HCS), the choice of $f_i^{(0)}$ in the driven case is a bit more intricate. This problem is also present of course in the monodisperse gas case \cite{GMT13,GCV13}. In some previous attempts \cite{G09}, the distributions $f_i^{(0)}$ were chosen to be \emph{stationary} at any point of the system. However, for general small deviations from the reference steady state, the collisional cooling cannot be compensated locally by the energy injected by the driving force in the system and so, $f_i^{(0)}$ is not in general a stationary distribution. As shown in previous studies for driven granular gases \cite{GMT13,GCV13,L06,G06}, the fact that $f_i^{(0)}$ is a time-dependent function introduces conceptual and practical difficulties not present when $f_i^{(0)}$ is assumed to be stationary \cite{G09}.

The irreversible parts of the mass, heat, and momentum fluxes are calculated here up to first order in the spatial gradients of the hydrodynamic fields. In addition, there is a new contribution (not present for dilute undriven mixtures) to the cooling rate proportional to the divergence of the flow velocity field. Therefore, as happens for freely cooling granular mixtures \cite{GD02}, the integral equations defining the transport coefficients for a driven binary mixture are somewhat more complicated than for the one-component driven case \cite{GCV13}: twelve coupled integral equations with nine transport coefficients. Thus, the explicit determination of the complete set of transport coefficients of the mixture is actually a very long task. For this reason, in this paper we will focus on the evaluation of the transport coefficients associated with the mass flux (four diffusion coefficients) and the shear viscosity coefficient.

One of the motivations of our study is to propose a kinetic equation that captures the influence of gas phase on the transport properties of grains through the action of nonconservative external forces. In fact, in the monodisperse case, our model reduces to a recent kinetic equation \cite{GTSH12} proposed to analyze several properties of gas-solid suspensions. In this context, we expect that our study has obvious applications in mesoscopic systems such as colloids and bidisperse suspensions \cite{J00,K90,KH01,BGP11}.

The plan of the paper is as follows. In Sec.\ \ref{sec2}, the coupled set of Boltzmann equations for the binary mixture and the corresponding hydrodynamic equations are recalled. Section \ref{sec3} analyzes the steady homogeneous state. As in the monodisperse case \cite{GMT12,CVV13}, scaling solutions $\varphi_{i,\text{s}}$ are proposed whose dependence on temperature $T$ and pressure $p$ occurs through two dimensionless parameters: the dimensionless velocity $\mathbf{c}=\mathbf{v}/v_0$ ($v_0$ being the thermal speed) and the reduced noise strength $\xi^*$. This contrasts with the results obtained in the HCS \cite{GD99} where $\varphi_{i,\text{s}}$ depends on $T$ and $p$ \emph{only} through $\mathbf{c}$.  Once the steady state is well characterized, in Sec.\ \ref{sec4} the CE expansion adapted to dissipative dynamics is used to construct the distribution functions to linear order in the gradients. This solution is used to evaluate the fluxes and identify the transport coefficients. As for elastic collisions, these coefficients are given in terms of the solutions of a set of coupled linear integral equations. A Sonine polynomial approximation is applied in Sec.\ \ref{sec5} to solve the integral equations defining the diffusion transport coefficients and the shear viscosity coefficient. These coefficients are explicitly determined as functions of the parameters of thermostat, the coefficients of restitution, and the masses, concentrations, and sizes of the constituents of the mixture. Comparisons with simulations carried out in the Brownian limit \cite{SVCP10} and with some previous theoretical results \cite{G09} obtained by using a \emph{local} stochastic thermostat are carried out in Sec.\ \ref{sec6}. The paper is closed in Sec.\ \ref{sec7} with a brief discussion of the results derived here.

\section{Bolztmann kinetic theory for driven granular binary mixtures}
\label{sec2}

We consider a granular binary mixture of inelastic hard spheres in $d$ dimensions with masses $m_i$ and diameters $\sigma_i$ ($i=1,2$). In the low-density regime, one can assume that there are no correlations between the velocities of two particles that are about to collide (molecular chaos hypothesis), so that the two-body distribution functions factorize into the product of the one-particle distribution functions $f_i(\mathbf{r},\mathbf{v},t)$. These distributions verify the set of nonlinear Boltzmann equations \cite{BDS97}
\begin{equation}
\partial_{t}f_i+\mathbf{v}\cdot \nabla f_i+{\cal F}_if_i=\sum_{j=1}^2\; J_{ij}[\mathbf{v}|f_i,f_j], \label{2.1}
\end{equation}
where the Boltzmann collision operator $J_{ij}[f_i,f_j]$ is
\begin{eqnarray}
\label{2.2}
J_{ij}\left[\mathbf{v}_1|f_i, f_j\right] &=& \sigma_{ij}^{d-1}\int
\dd\mathbf{v}_{2}\int \dd\widehat{\boldsymbol {\sigma}}\Theta
(\widehat{\boldsymbol {\sigma}}\cdot \mathbf{g}_{12})(\widehat{
\boldsymbol {\sigma }}\cdot \mathbf{g}_{12})\nonumber\\
& & \times\left[ \alpha_{ij}^{-2}f_i(\mathbf{r},\mathbf{v}_{1}^{\prime
},t)f_j(\mathbf{r},\mathbf{v}_{2}^{\prime
},t)\right.\nonumber\\
& & \left.-f_i(\mathbf{r},\mathbf{v}_{1},t)f_j(\mathbf{r},\mathbf{v}_{2},t)\right].
\end{eqnarray}
Here, $\sigma_{ij}=(\sigma_i+\sigma_j)/2$, $\widehat{\boldsymbol {\sigma}}$ is a unit vector directed along the line of centers from the sphere of species $i$ to that of species $j$ at
contact, $\Theta$ is the Heaviside step function, and $\mathbf{g}_{12}=\mathbf{v}_1-\mathbf{v}_2$ is the relative velocity. The precollisional velocities are
\begin{equation}
\label{2.3.1}
\mathbf{ v}_{1}^{\prime }=\mathbf{ v}_{1}-\mu _{ji}\left( 1+\alpha
_{ij}^{-1}\right) (\widehat{{\boldsymbol {\sigma }}}\cdot {\bf
g}_{12})\widehat{{\boldsymbol {\sigma }}} ,\nonumber\\
\eeq
\beq
\mathbf{ v}_{2}^{\prime }=\mathbf{ v}_{2}+\mu _{ij}\left( 1+\alpha
_{ij}^{-1}\right) (\widehat{{\boldsymbol {\sigma }}}\cdot {\bf
g}_{12})\widehat{ \boldsymbol {\sigma}} ,  \label{2.3.2}
\end{equation}
where $\mu _{ij}= m_{i}/\left( m_{i}+m_{j}\right)$ and $\alpha_{ij}\leq 1$ is the (constant) coefficient of normal restitution for collisions $(i,j)$. Moreover, in Eq.\ \eqref{2.1} ${\cal F}_i$ is an operator representing the effect of an external force.

In order to maintain a fluidized granular mixture, an external energy source is needed to compensate for the collisional cooling. As said in the Introduction, it is quite usual in computer simulations to homogeneously heat the system by means of an external driving force (thermostat). Here, as in our previous work \cite{GCV13} for monodisperse granular gases, we will assume that the external force is composed by two independent terms. One term corresponds to a drag force ($\mathbf{F}^{\text{drag}}$) proportional to the velocity of the particle. The other term corresponds to a stochastic force ($\mathbf{F}^{\text{st}}$) where the particles are randomly kicked between collisions \cite{WM96}. As usual, the stochastic force is assumed to have the form of a Gaussian white noise and is represented by a Fokker-Planck collision operator of the form $\partial^2 f_i/\partial v^2$  in the Boltzmann equation \cite{NE98}. While the term $\mathbf{F}^{\text{drag}}$ mimics the effect of the interstitial gas phase, the noise force $\mathbf{F}^{\text{st}}$ tries to simulate the kinetic energy gain due to eventual collisions with the (more rapid) particles of the surrounding fluid. This type of thermostat composed by two terms has been widely used by Puglisi and coworkers \cite{puglisi} in several previous works.

On the other hand, there is some flexibility in the choice of the explicit forms of $\mathbf{F}^{\text{drag}}$ and $\mathbf{F}^{\text{st}}$  for multicomponent systems since either one takes both forces to be the same for each species \cite{HBB00,BT02,DHGD02} or they can be chosen to be functions of the mass of each species \cite{puglisi}. To cover both possibilities, we will assume that the drag and stochastic forces contribute to the Boltzmann equation \eqref{2.1} with terms of the form
\beq
\label{2.4}
\mathcal{F}_if_i=\mathcal{F}_i^{\text{drag}}f_i+\mathcal{F}_i^{\text{st}}f_i,
\eeq
where
\beq
\label{2.5}
\mathcal{F}_i^{\text{drag}}f_i=-\frac{\gamma_\text{b}}{m_i^{\beta}}
\frac{\partial}{\partial\mathbf{v}}\cdot\left(\mathbf{v}-\mathbf{U}_\text{g}\right)f_i,
\eeq
\beq
\label{2.6}
\mathcal{F}_i^{\text{st}}f_i=-\frac{1}{2}\frac{\xi_\text{b}^2}{m_i^\lambda}\frac{\partial^2}{\partial v^2} f_i.
\eeq
In Eqs.\ \eqref{2.5}--\eqref{2.6}, $\beta$ and $\lambda$ are \emph{arbitrary} constants of the driven model, $\gamma_\text{b}$ is the drag (or friction) coefficient, and $\xi_\text{b}^2$ represents the strength of the correlation in the Gaussian white noise. In addition, since our model pretends to incorporate the effect of gas phase into the dynamics of grains, in Eq.\ \eqref{2.5} we have considered the ``peculiar'' velocity $\mathbf{v}-\mathbf{U}_\text{g}$ (rather than the instantaneous velocity $\mathbf{v}$ of particle) in the drag force expression. Here, $\mathbf{U}_\text{g}$ can be interpreted as the mean velocity of gas surrounding the solid particles and is assumed to be a known quantity of the model. The parameters $\beta$ and $\lambda$ can be seen as free parameters of the model. In particular, when $\gamma_\text{b}=0$ and $\lambda=0$ our thermostat reduces to the stochastic thermostat used in previous works \cite{BT02,DHGD02} for granular mixtures while the choice $\beta=1$ and $\lambda=2$ reduces to the conventional Fokker-Planck model for ordinary (elastic) mixtures \cite{puglisi,H03}. This latter version of the model has been also used to analyze granular Brownian motion \cite{SVCP10}. Thus, our model can be seen as a generalization of previous driven models and only specific values of $\beta$ and $\lambda$ will be considered at the end of the calculations to make contact with some particular situations \cite{G09}.

The Boltzmann kinetic equation \eqref{2.1} can be more explicitly written when one takes into account the form \eqref{2.4} of the forcing term $\mathcal{F}_if_i$. It can be written as
\beqa
\partial_{t}f_i&+&\mathbf{v}\cdot \nabla f_i-\frac{\gamma_\text{b}}{m_i^{\beta}}\Delta \mathbf{U} \cdot
\frac{\partial}{\partial\mathbf{v}}f_i-\frac{\gamma_\text{b}}{m_i^{\beta}}
\frac{\partial}{\partial\mathbf{v}}\cdot \mathbf{V}
f_i\nonumber\\
& & -\frac{1}{2}\frac{\xi_\text{b}^2}{m_i^{\lambda}}\frac{\partial^2}{\partial v^2}f_i=\sum_{j=1}^2\; J_{ij}[f_i,f_j], \label{2.7}
\eeqa
where $\Delta \mathbf{U}=\mathbf{U}-\mathbf{U}_\text{g}$ and $\mathbf{V}(\mathbf{r},t)=\mathbf{v}-\mathbf{U}(\mathbf{r},t)$. Here, $\mathbf{U}(\mathbf{r},t)$ is the mean flow velocity of grains defined as
\beq
\label{2.8}
\rho \mathbf{U}=\sum_{i=1}^2\int\dd \mathbf{ v}m_{i}\mathbf{v}f_{i}(\mathbf{ v}),
\eeq
where $\rho=\sum_{i=1}^2 m_in_i$ is the total mass density. In addition,
\beq
\label{2.9}
n_{i}=\int \dd\mathbf{ v}\;f_{i}(\mathbf{ v}),
\eeq
is the local number density of species $i$. It is important to remark that in the case of a monodisperse granular gas (for $\beta=1$ and $\lambda=0$), the Boltzmann equation \eqref{2.7} is similar to the one recently proposed \cite{GTSH12} to model the effects of the interstitial fluid on grains in monodisperse gas-solid suspensions.

Apart from the fields $n_i$ and $\mathbf{U}$, the other relevant hydrodynamic field of the mixture is the granular temperature $T(\mathbf{r},t)$. It is defined as
\beq
\label{2.10}
T=\frac{1}{n}\sum_{i=1}^2\int \dd\mathbf{v}\frac{m_{i}}{d}V^{2}f_{i}(\mathbf{ v})\;,
\eeq
where $n=n_{1}+n_{2}$ is the total number density. At a kinetic level, it is also convenient to introduce the partial kinetic temperatures $T_i$ for each species defined as
\begin{equation}
\label{2.11}
T_i=\frac{m_{i}}{d n_i}\int\; \dd\mathbf{ v}\;V^{2}f_{i}(\mathbf{ v}).
\end{equation}
The partial temperatures $T_i$ measure the mean kinetic energy of each species. According to Eq.\ \eqref{2.10}, the granular temperature $T$ of the mixture can be also written as
\beq
\label{2.12}
T=\sum_{i=1}^2\, x_i T_i,
\eeq
where $x_i=n_i/n$ is the mole fraction of species $i$.

The collision operators conserve the particle number of each
species and the total momentum but the total energy is not
conserved:
\begin{equation}
\int \dd\mathbf{ v}J_{ij}[\mathbf{ v}|f_{i},f_{j}]=0,  \label{2.13}
\end{equation}
\begin{equation}
\sum_{i=1}^2\sum_{j=1}^2m_i\int \dd\mathbf{ v}\;\mathbf{ v}
J_{ij}[{\bf v}|f_{i},f_{j}]=0, \label{2.14}
\end{equation}
\begin{equation}
\sum_{i=1}^2\sum_{j=1}^2m_i\int \dd\mathbf{ v}\; V^{2}J_{ij}[\mathbf{ v}
|f_{i},f_{j}]=-d nT\zeta,  \label{2.15}
\end{equation}
where  $\zeta$ is identified as the total ``cooling rate'' due to inelastic collisions among all species. The corresponding partial ``cooling rates'' $\zeta_i$ for the partial temperatures $T_i$ are defined as
\begin{equation}
\label{2.16} \zeta_i=\sum_{j=1}^2\zeta_{ij}=-
\frac{m_i}{dn_iT_i}\sum_{j=1}^2\int \dd\mathbf{ v}\; V^{2}J_{ij}[{\bf
v}|f_{i},f_{j}],
\end{equation}
where the second equality defines the quantities $\zeta_{ij}$. The total cooling rate $\zeta$ can be written in terms of the partial cooling rates $\zeta_i$ as
\begin{equation}
\label{2.17} \zeta=T^{-1}\sum_{i=1}^2\;x_iT_i\zeta_i.
\end{equation}

From Eq. \eqref{2.7} and Eqs.\ (\ref{2.13})--(\ref{2.15}), the macroscopic balance equations for the mixture can be obtained. They are given by
\begin{equation}
D_{t}n_{i}+n_{i}\nabla \cdot \mathbf{U}+\frac{\nabla \cdot \mathbf{ j}_{i}}{m_{i}}
=0,  \label{2.18}
\end{equation}
\begin{equation}
D_{t}\mathbf{U}+\rho ^{-1}\nabla
\cdot\mathsf{P}=-\frac{\gamma_\text{b}}{\rho}\left(\Delta \mathbf{U}\sum_{i=1}^2\frac{\rho_i}{m_i^\beta}
+\sum_{i=1}^2\frac{\mathbf{ j}_{i}}{m_{i}^\beta}\right),  \label{2.19}
\end{equation}
\beqa
\label{2.20}
D_{t}T&-&\frac{T}{n}\sum_{i=1}^2\frac{\nabla \cdot \mathbf{ j}_{i}}{m_{i}}+\frac{2}{dn}
\left( \nabla \cdot \mathbf{ q}+{\sf P}:\nabla \mathbf{U}\right) =\nonumber\\
& & -\frac{2 \gamma_\text{b}}{d n} \sum_{i=1}^2\frac{
\Delta \mathbf{U} \cdot\mathbf{ j}_{i}}{m_{i}^\beta}-2\gamma_\text{b}\sum_{i=1}^2
\frac{x_i T_i}{m_i^\beta}\nonumber\\
& & +\frac{\xi_\text{b}^2}{n}\sum_{i=1}^2\frac{\rho_i}{m_i^\lambda}-\zeta \,T.
\eeqa
In the above equations, $D_{t}=\partial _{t}+\mathbf{U}\cdot \nabla $ is the
material derivative, $\rho_i=m_in_i$,
\begin{equation}
\mathbf{j}_{i}=m_{i}\int \dd\mathbf{ v}\,\mathbf{V}\,f_{i}(\mathbf{ v})\;,
\label{2.21}
\end{equation}
is the mass flux for species $i$ relative to the local flow
\begin{equation}
\mathsf{P}=\sum_{i=1}^2\,\int \dd\mathbf{ v}\,m_{i}\mathbf{V}\mathbf{V}\,f_{i}(\mathbf{ v}),  \label{2.22}
\end{equation}
is the total pressure tensor, and
\begin{equation}
\mathbf{ q}=\sum_{i=1}^2\,\int d\mathbf{ v}\,\case{1}{2}m_{i}V^{2}\mathbf{V}\,f_{i}(\mathbf{ v}),  \label{2.23}
\end{equation}
is the total heat flux. Note that $\mathbf{j}_1=-\mathbf{j}_2$ by definition of the flow velocity $\textbf{U}$.

The balance equations  \eqref{2.18}-\eqref{2.20} become a closed set of hydrodynamic equations for the
fields $n_{i}$, $\mathbf{U}$ and $T$ once the fluxes (\ref{2.21})--(\ref{2.23})
and the cooling rate \eqref{2.15} are obtained in terms of the hydrodynamic
fields and their gradients. The resulting equations constitute the
hydrodynamics for the driven mixture. Since these fluxes are explicit linear
functionals of $f_{i}$, a representation in terms of the hydrodynamic fields results
when a solution to the Boltzmann equation can be obtained as a function of
the fields and their gradients. Such a solution is called a \emph{normal} or hydrodynamic
solution and can be obtained for small spatial gradients from the Chapman-Enskog method \cite{CC70}. This solution will be worked out in Sec.\ \ref{sec4}.

\section{Homogeneous steady states}
\label{sec3}

Before considering inhomogeneous problems, it is quite instructive to study first the homogeneous state. In this situation, the partial densities $n_i(\mathbf{r},t)= n_{i,\text{s}}$ are constant, the granular temperature $T(\mathbf{r},t)= T(t)$ is spatially uniform, and, with an appropriate selection of the frame of reference, the mean flow velocities vanish ($\mathbf{U}=\mathbf{U}_\text{g}=\mathbf{0}$). Under these conditions, Eq.\ \eqref{2.7} for $f_i(\mathbf{v},t)$ becomes
\beq
\partial_{t}f_i-\frac{\gamma_\text{b}}{m_i^{\beta}}
\frac{\partial}{\partial\mathbf{v}}\cdot \mathbf{v}
f_i-\frac{1}{2}\frac{\xi_\text{b}^2}{m_i^{\lambda}}\frac{\partial^2}{\partial v^2}f_i=\sum_{j=1}^2\; J_{ij}[f_i,f_j]. \label{3.1}
\eeq
The balance equation \eqref{2.20} for the temperature reads simply
\begin{equation}
\label{3.2}
\partial_tT=-2\gamma_\text{b}\sum_{i=1}^2
\frac{x_i T_i}{m_i^\beta}+\frac{\xi_\text{b}^2}{n}\sum_{i=1}^2\frac{\rho_i}{m_i^\lambda}-\zeta T.
\end{equation}
Analogously, the evolution equation for the partial temperatures $T_i$ can be obtained by multiplying both sides of Eq.\ \eqref{3.1} by $\frac{m_i}{2}V^2$ and integrating over $\mathbf{v}$. The result is
\begin{equation}
\label{3.3}
\partial_tT_i=-\frac{2 T_i}{m_i^\beta}\gamma_\text{b} +\frac{\xi_\text{b}^2}{m_i^{\lambda-1}}-\zeta_i T_i.
\end{equation}

As said before, we are here only interested in the normal solution to Eq.\ \eqref{3.1}. In this case, the distribution function $f_i$ depends on time only through the temperature $T$ \cite{GD99}:
\beqa
\label{3.4}
\partial_t f_i=\frac{\partial f_i}{\partial T}\partial_tT &=&-\left(2\gamma_\text{b}\sum_{i=1}^2
\frac{x_i \chi_i}{m_i^\beta}-\frac{\xi_\text{b}^2}{p}\sum_{i=1}^2\frac{\rho_i}{m_i^\lambda}+\zeta\right)\nonumber\\
& \times&
T\frac{\partial f_i}{\partial T},
\eeqa
where $\chi_i= T_i/T$ is the temperature ratio for species $i$. As widely discussed in the free cooling case \cite{GD99}, the fact that $f_i$ qualifies as normal solution implies necessarily that the temperature ratios $\chi_i$ are independent of time but different from one for inelastic collisions (breakdown of energy equipartition). The violation of equipartition theorem for granular mixtures has been extensively confirmed by computer simulations \cite{BT02,DHGD02,computer}, experiments \cite{exp} and kinetic theory calculations for undriven \cite{GD99} and driven \cite{BT02} systems.

After a transient regime, the system is expected to achieve a \emph{steady} state characterized by constant partial temperatures $T_{i,\text{s}}$. Thus, according to Eq.\ \eqref{3.3}, the (asymptotic) steady partial temperatures $T_{i,\text{s}}$ are given by
\begin{equation}
\label{3.5}
\frac{2 T_{i,\text{s}}}{m_i^\beta}\gamma_\text{b} +\zeta_{i,\text{s}} \,T_{i,\text{s}}=\frac{\xi_\text{b}^2}{m_i^{\lambda-1}},
\end{equation}
where the subindex $\text{s}$ means that the quantities are evaluated in the steady state.

In the case of elastic collisions ($\alpha_{ij}=1$) and if the distributions $f_{i,\text{s}}$ are Maxwellians at the same temperature, then $\zeta_i=0$ and Eq.\ \eqref{3.5} yields
\beq
\label{3.6}
T_{i,\text{s}}^{\text{el}}=\frac{\xi_\text{b}^2}{2\gamma_\text{b}m_i^{\lambda-\beta-1}}.
\eeq
According to Eq.\ \eqref{3.6}, the energy equipartition is fulfilled ($T_{1,\text{s}}=T_{2,\text{s}}$) if $m_1=m_2$ (for any choice of $\lambda$ and $\beta$) or $\lambda-\beta=1$ (for $m_1\neq m_2$). Therefore,
\beq
\label{3.7}
T_{1,\text{s}}^{\text{el}}=T_{2,\text{s}}^{\text{el}}=T_\text{b}=
\frac{\xi_\text{b}^2}{2\gamma_\text{b} (2\overline{m})^{\lambda-\beta-1}},
\eeq
where
\begin{equation}
\label{3.12}
\overline{m}=\frac{m_1m_2}{m_1+m_2}.
\end{equation}
Equation \eqref{3.7} defines a ``bath temperature'' $T_\text{b}$. Its name may be justified since it is determined by the two thermostat parameters ($\gamma_\text{b}$ and $\xi_\text{b}^2$) and it can be considered as remnant of the temperature of the surrounding elastic fluid. It is quite apparent that in general we find energy non-equipartition ($T_{1,\text{s}}^{\text{el}}\neq T_{2,\text{s}}^{\text{el}}$) even for elastic collisions when $\lambda-\beta \neq 1$. The condition $\lambda-\beta= 1$ to have energy equipartition in the elastic case should have been expected due to the definition of thermostat. Indeed it seems equivalent to the so-called ``fluctuation-dissipation relation of the second kind'' \cite{KTH85}.

In order to determine $T_{i,\text{s}}$ one has to obtain the steady state solution $f_{i,\text{s}}(\mathbf{v})$ to Eq.\ \eqref{3.1}. By using the relation \eqref{3.5}, in the steady state ($\partial_tf_i=0$) Eq.\ \eqref{3.1} becomes
\beqa
\frac{1}{2}\zeta_{i,\text{s}}
\frac{\partial}{\partial{\bf v}}\cdot {\bf v}
f_{i,\text{s}}&-&\frac{1}{2}\frac{\xi_\text{b}^2}{m_i^{\lambda-1}T_{i,\text{s}}}
\frac{\partial}{\partial{\bf v}}\cdot {\bf v}
f_{i,\text{s}}-\frac{1}{2}\frac{\xi_\text{b}^2}{m_i^{\lambda}}\frac{\partial^2}{\partial v^2}f_{i,\text{s}}\nonumber\\
&=&\sum_{j=1}^2\; J_{ij}[\mathbf{v}|f_{i,\text{s}},f_{j,\text{s}}]. \label{3.8}
\eeqa
As in the monocomponent case \cite{GCV13}, it is expected that $f_{i,\text{s}}$ depends on the model parameters $\gamma_\text{b}$ and $\xi_\text{b}^2$. Although the explicit form of $f_{i,\text{s}}$ is not known, dimensional analysis requires that $f_{i,\text{s}}$ has the scaled form
\begin{equation}
\label{3.9} f_{i,\text{s}}({\bf v},\gamma_\text{b}, \xi_\text{b}^2)=n_{i,\text{s}}v_0^{-d}\varphi_{i,\text{s}}\left(x_1,\mathbf{c}, \xi_\text{s}^*, \gamma_\text{s}^* \right),
\end{equation}
where $\varphi_{i,\text{s}}$ is an unknown function of the dimensionless  parameters
\begin{equation}
\label{3.10}
\mathbf{c}= \frac{\mathbf {v}}{v_0}, \quad \xi_\text{s}^*= \frac{\xi_\text{b}^2}{n_{\text{s}}\sigma_{12}^{d-1}\overline{m}^{\lambda-1} T_\text{s}v_0},
\end{equation}
and
\beq
\label{3.10.1}
\gamma_\text{s}^*= \frac{\gamma_\text{b}}{n_{\text{s}}\sigma_{12}^{d-1}\overline{m}^{\beta}v_0}.
\eeq
Here, $T_\text{s}=x_1T_{1,\text{s}}+x_2T_{2,\text{s}}$ is the steady value of the granular temperature and $v_0=\sqrt{2T_\text{s}/\overline{m}}$ is the thermal speed. The (reduced) drag parameter $\gamma_\text{s}^*$ can be easily expressed in terms of the (reduced) noise strength $\xi_\text{s}^*$ and density as
\beq
\label{4.10}
\gamma_\text{s}^*=\omega_\text{s}^* \xi_\text{s}^{*1/3}, \quad
\omega_\text{s}^*= \frac{\gamma_\text{b}}{\overline{m}^\beta} \left(\frac{\overline{m}^\lambda}{2 \xi^2_b}\right)^{1/3} \left(n_\text{s}\sigma_{12}^{d-1}\right)^{-2/3}.
\eeq
Note that, when Eq.\ \eqref{4.10} is used, the dependence of the scaled distribution function $\varphi_{i,\text{s}}$ on temperature is encoded through two parameters: the dimensionless velocity $\mathbf{c}$ and the (reduced) noise strength $\xi_\text{s}^*$. This scaling differs from the one assumed in the case of the free cooling case \cite{GD99} where only the dimensionless velocity $\mathbf{c}$ is required to characterize the temperature dependence of the scaled distributions $\varphi_{i,\text{s}}$.

In terms of the (reduced) distribution function $\varphi_{i,\text{s}}$, Eq.\ \eqref{3.8} can be rewritten as
\beqa
\label{3.13}
\frac{1}{2}\zeta_{i,\text{s}}^* \frac{\partial}{\partial {\bf
c}}\cdot {\bf c}\varphi_{i,\text{s}}&-&\frac{1}{2}\frac{\xi_\text{s}^*}{M_i^{\lambda-1}\chi_i}
\frac{\partial}{\partial {\bf c}}\cdot {\bf c} \varphi_{i,\text{s}}- \frac{1}{4}\frac{\xi_\text{s}^*}{M_i^\lambda} \frac{\partial^2}{\partial c^2}\varphi_{i,\text{s}}\nonumber\\
&=&
\sum_{j=1}^2\; J_{ij}^*[\mathbf{c}|\varphi_{i,\text{s}},\varphi_{j,\text{s}}],
\eeqa
where $M_i= m_i/\overline{m}$, $\chi_{i,\text{s}}= T_{i,\text{s}}/T_{\text{s}}$,
\begin{equation}
\label{3.14}
\zeta_{i,\text{s}}^*=\frac{\zeta_{i,\text{s}}}{n_\text{s}\sigma_{12}^{d-1}v_0},
\end{equation}
and
\begin{eqnarray}
\label{3.15}
& &J_{ij}^*[\mathbf{c}|\varphi_{i,\text{s}},\varphi_{j,\text{s}}]= \frac{J_{ij}[\mathbf{v}|f_{i,\text{s}},f_{j,\text{s}}]} {n_\text{s}\sigma_{12}^{d-1}n_{i,\text{s}}v_0^{1-d}}\nonumber\\
&=&x_{j,\text{s}} \left(\frac{\sigma_{ij}}{\sigma_{12}}\right)^{d-1}\int
\dd\mathbf{c}_{2}\int \dd\widehat{\boldsymbol {\sigma}}\Theta
(\widehat{\boldsymbol {\sigma}}\cdot \mathbf{g}_{12}^*)(\widehat{
\boldsymbol {\sigma }}\cdot \mathbf{g}_{12}^*)\nonumber\\
&\times &\left[ \alpha_{ij}^{-2}\varphi_{i,\text{s}}(\mathbf{c}_{1}^{\prime
})\varphi_{j,\text{s}}(\mathbf{c}_{2}^{\prime})-\varphi_{i,\text{s}}
(\mathbf{c}_{1})\varphi_{j,\text{s}}(\mathbf{c}_{2})\right].\nonumber\\
\end{eqnarray}
Here, $x_{i,\text{s}}=n_{i,\text{s}}/n_{\text{s}}$ and $\mathbf{g}_{12}^*=\mathbf{c}_1-\mathbf{c}_2$. Similarly, in dimensionless variables the cooling rates are given by
\begin{equation}
\label{3.16}
\zeta_{i,\text{s}}^*=-\frac{2}{d}\frac{M_i}{\chi_{i,\text{s}}}\sum_{j=1}^2\;\int\dd\mathbf{c}\; c^2\; J_{ij}^*[\varphi_{i,\text{s}},\varphi_{j,\text{s}}].
\end{equation}

The (reduced) partial temperatures $T_{i,\text{s}}^*= T_{i,\text{s}}/T_\text{b}$ can be determined from the condition \eqref{3.5} for $i=1,2$. The corresponding equations can be written as
\beq
\label{3.17}
T_{\text{s}}^{*}\left[1-(M_i/2)^{\lambda-1-\beta}T_{i,\text{s}}^*\right]\xi_\text{s}^*
=M_i^{\lambda-1}\zeta_{i,\text{s}}^* T_{i,\text{s}}^*,
\eeq
where $T_{\text{s}}^*= T_\text{s}/T_\text{b}$.


Once the reduced distributions $\varphi_{1,\text{s}}$ and $\varphi_{2,\text{s}}$ have been obtained from Eqs.\ \eqref{3.13}, the integrals on the right-hand side of Eq.\ \eqref{3.16} can be performed to determine the partial cooling rates $\zeta_{i,\text{s}}^*$. Then, the partial temperatures $T_{i,\text{s}}^*$ can be finally obtained from Eqs.\ \eqref{3.17} (for $i=1,2$) in terms of the model parameters $T_\text{b}$ and $\xi_\text{b}^2$, the concentration $x_1$ and the mechanical parameters of the mixture (masses, diameters, and coefficients of restitution).

As said before, the exact form of the distributions $\varphi_{i,\text{s}}$ is not known. However, previous results derived for driven granular mixtures \cite{BT02,G09} have shown that a good estimate for the partial temperatures can be obtained by using Maxwellians at different temperatures for $\varphi_{i,\text{s}}(\mathbf{c})$:
\beq
\label{3.19}
\varphi_{i,\text{s}}(\mathbf{c})\to \varphi_{i,\text{M}}(\mathbf{c})=\pi^{-d/2} \theta_{i}^{d/2}\; e^{-\theta_{i} c^2},
\end{equation}
where $\theta_{i}=M_i/\chi_{i,\text{s}}$. With this approximation, one gets \cite{GD99}
\begin{eqnarray}
\label{3.20}
\zeta_{i,\text{s}}^*&=&\frac{4\pi^{(d-1)/2}}{d\Gamma\left(\frac{d}{2}\right)}
\sum_{j=1}^2
x_{j,\text{s}}\mu_{ji}\left(\frac{\sigma_{ij}}{\sigma_{12}}\right)^{d-1}\left(\frac{\theta_i+\theta_j}
{\theta_i\theta_j}\right)^{1/2}\nonumber\\
& &\times (1+\alpha_{ij})
\left[1-\frac{\mu_{ji}}{2}(1+\alpha_{ij})
\frac{\theta_i+\theta_j}{\theta_j}\right].
\end{eqnarray}
Substitution of Eq.\ \eqref{3.20} into Eqs.\ \eqref{3.17} allows us to get the partial temperatures $T_{i,\text{s}}^*$.

An interesting limit situation corresponds to granular Brownian motion, namely, when the mass of the tracer species ($x_1\to 0$) is much heavier than the particles of the excess granular gas ($m_1\gg m_2$). In this limit case, $M_2 \to 1$, $M_1 \to m_1/m_2$ and the tracer temperature $T_{1,\text{s}}$ is given by
\begin{equation}
\label{3.21}
T_{1,\text{s}}=\frac{\left(\frac{2m_2}{m_1}\right)^{\lambda-\beta-1}\gamma_\text{b}T_\text{b}+\gamma_\text{g} \frac{1+\alpha_{12}}{2}T_{2,\text{s}}}{\gamma_\text{b}+\gamma_\text{g}}.
\end{equation}
Here,
\beq
\label{3.22}
\gamma_\text{g}=\frac{2\pi^{(d-1)/2}}{d\Gamma\left(\frac{d}{2}\right)}
(1+\alpha_{12})m_1^{\beta-1}m_2n_2\sigma_{12}^{d-1}\sqrt{\frac{2T_{2,\text{s}}}{m_2}},
\eeq
and the temperature of granular gas $T_{2,\text{s}}$ obeys the equation
\begin{equation}
\label{3.23}
T_{2,\text{s}}=2^{\lambda-\beta-1}T_\text{b}-\frac{\pi^{(d-1)/2}}
{d\Gamma\left(\frac{d}{2}\right)}\frac{n_2\sigma_2^{d-1}(1-\alpha_{22}^2)} {\gamma_\text{b}}
m_2^{\beta-\frac{1}{2}}T_{2,\text{s}}^{3/2}.
\end{equation}
In the two-dimensional case ($d=2$), Eqs.\ \eqref{3.20}--\eqref{3.23} agree with the results derived by Sarracino \emph{et. al} \cite{SVCP10} for hard disks when $\beta=1$ and $\lambda=2$.

\begin{figure}
  \includegraphics[width=.45\textwidth]{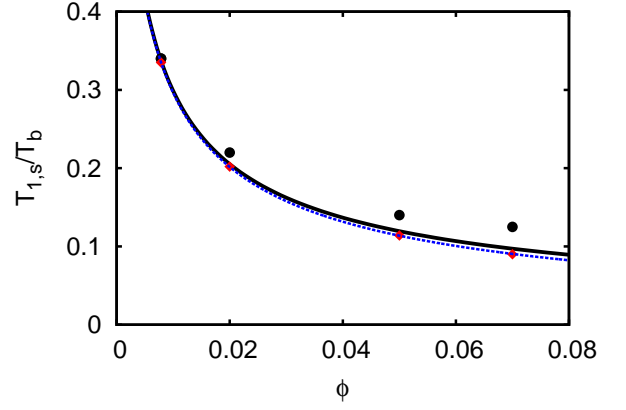}
\caption{(Color online) Plot of the (steady) reduced temperature $T_{1,\text{s}}/T_\text{b}$ of a Brownian particle as a function of the volume fraction $\phi$ for hard disks ($d=2$). The parameters of the system (impurity particle plus granular gas) are $m_1=100m_2$, $\sigma_1=\sigma_2$, and $\alpha_{11}=\alpha_{12}=\alpha_{22}=0.8$. The solid line refers to the results derived from Eq.\ \eqref{3.17} while the dashed line corresponds to the results obtained from Eq. \eqref{3.21} in the Brownian limit ($m_1/m_2\to \infty$). In both cases, $\beta=1$ and $\lambda=2$.
Symbols are the simulation results obtained in Ref. \cite{SVCP10} by means of DSMC method (red diamonds) and MD simulations (black circles).}
\label{fig1}
\end{figure}
\begin{figure}
  \includegraphics[width=.45\textwidth]{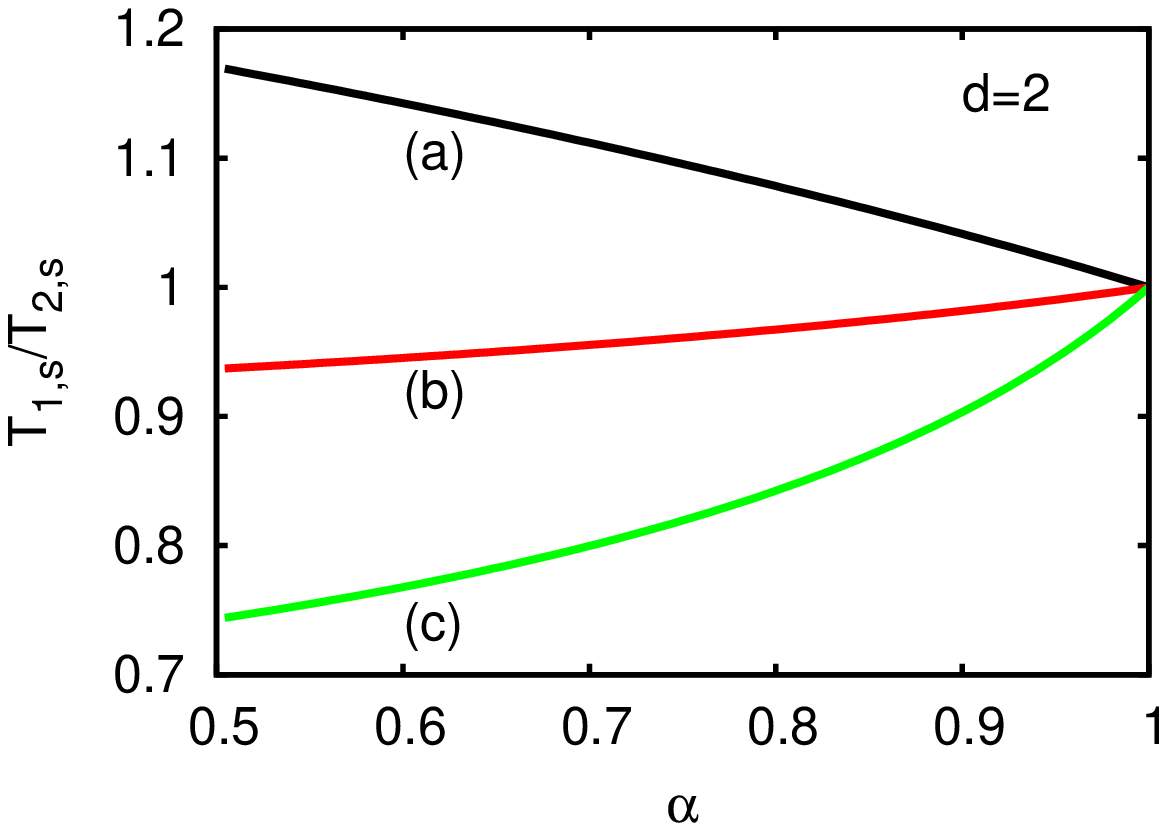}
  \includegraphics[width=.45\textwidth]{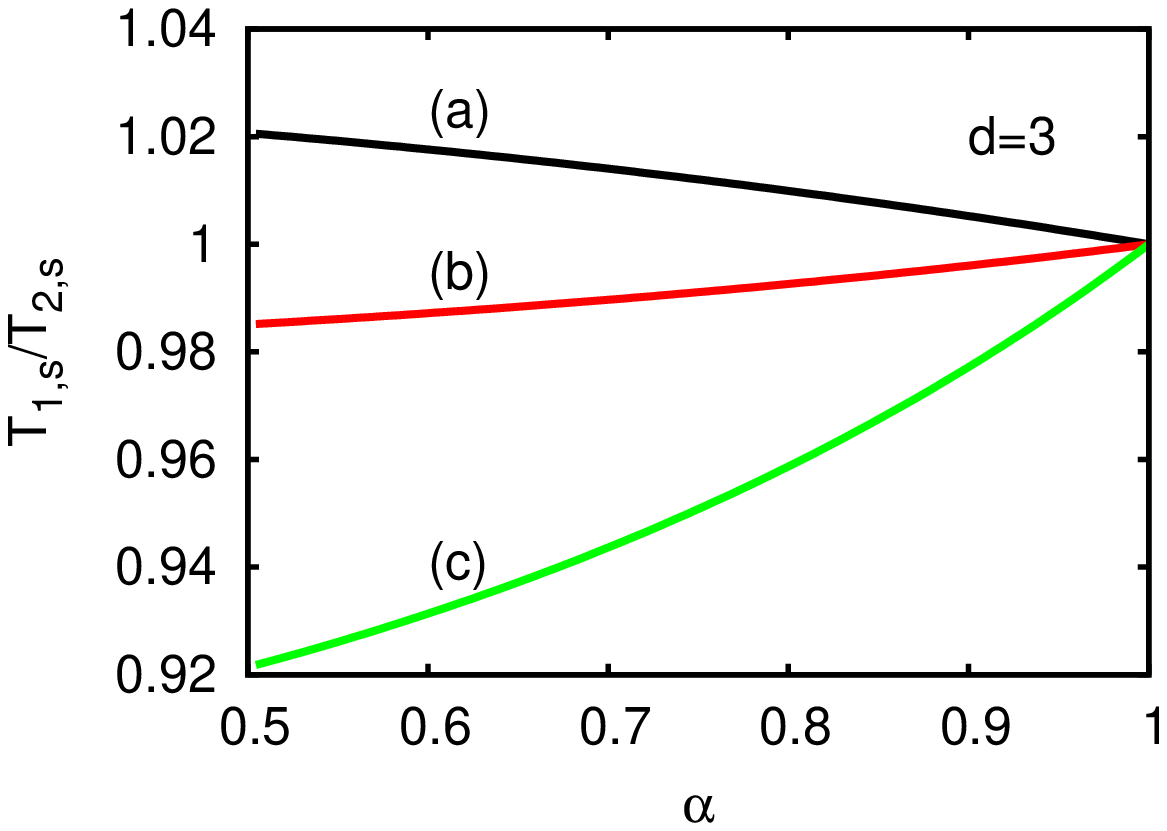}
\caption{(Color online) Temperature ratio $T_{1,\text{s}}/T_{2,\text{s}}$ versus the (common) coefficient of restitution
$\alpha$ for hard disks (top panel) and hard spheres (bottom panel) for $x_1=\frac{2}{3}$, $\sigma_1=\sigma_2$ and three different values of the mass ratio $m_1/m_2$: (a) $m_1/m_2=0.1$, (b) $m_1/m_2=2$, and (c) $m_1/m_2=10$. The parameters of the system are the same as those considered in Fig.\ \ref{fig1}}
\label{fig1bis}
\end{figure}

Figure \ref{fig1} shows the (steady) reduced temperature $T_{1,\text{s}}/T_\text{b}$  versus the volume fraction $\phi=\pi n_2 \sigma_{2}^2/4$ of the excess gas in the tracer limit ($x_1\to 0$) for the case $m_1=100m_2$, $\sigma_1=\sigma_2$, and $\alpha_{11}=\alpha_{22}=\alpha_{12}=0.8$. The theoretical results derived from Eqs.\ \eqref{3.17} and \eqref{3.21} (Brownian limit, $m_1/m_2\to \infty$) for hard disks ($d=2$) are compared with those obtained in Ref.\ \cite{SVCP10} by means of molecular dynamics simulations (MD) and by numerically solving the Langevin equation from the direct simulation Monte Carlo (DSMC) method \cite{B94}. As in Ref.\ \cite{SVCP10}, $\beta=1$ and $\lambda=2$ and the fixed parameters of the simulations are $m_2=1$, $\sigma_2=0.01$, $\gamma_\text{b}=0.1$, and  $\xi_\text{b}^2=0.2$. This gives a bath temperature $T_\text{b}=1$. We observe a good agreement between both theories and simulations in the complete range of values of $\phi$ considered. Given that the DSMC method numerically solves the Langevin equation (which is obtained from the Boltzmann equation in the limit $m_1/m_2\to \infty$), the theoretical predictions obtained from Eq.\ \eqref{3.21} compares slightly better with DSMC results than those derived from Eq.\ \eqref{3.17} (which are obtained for the mass ratio $m_1/m_2=100$). On the other hand, as expected, MD simulations are closer to the results derived from Eq.\ \eqref{3.17} than those obtained from Eq.\ \eqref{3.21}.

The dependence of the temperature ratio $T_{1,\text{s}}/T_{2,\text{s}}$ on the (common) coefficient of restitution $\alpha_{11}=\alpha_{22}=\alpha_{12}\equiv \alpha$ is shown in Fig.\ \ref{fig1bis} for hard disks ($d=2$) and spheres ($d=3$). We have considered a binary mixture where $x_1=\frac{2}{3}$, $\sigma_1/\sigma_2$=1, and three different values of the mass ratio $m_1/m_2$. The values of the parameters of the system are the same as those considered before in Fig.\ \ref{fig1}. We observe that the deviations from the energy equipartition ($T_{1,\text{s}}=T_{2,\text{s}}$) are smaller than those previously reported for undriven granular mixtures \cite{GD99}. Moreover, in contrast to the free cooling case, the energy of the lighter particle is larger than that of the heavier particle. This means that the impact of thermostat on the temperature ratio is significant since the qualitative behavior of the latter on the mass ratio is the opposite as the one found in the undriven case.

\section{Chapman-Enskog solution of the Boltzmann equations}
\label{sec4}

The homogeneous steady state analyzed in Sec.\ \ref{sec3} can be disturbed by the presence of small spatial gradients. These gradients give rise to nonzero contributions to the mass, momentum, and heat fluxes, which are characterized by transport coefficients. The determination of the transport coefficients of the mixture is the main goal of the present paper. However, as pointed out in the Introduction, the study of transport in multicomponent systems is more intricate than for monocomponent systems not only from a fundamental point of view (for instance, there are cross transport effects not present in single gases) but also from a more practical point of view (there are more coupled integral equations to solve than in single gases).

As in our previous effort for driven monodisperse gases \cite{GCV13}, we consider states that deviate from steady homogeneous states by \emph{small} spatial gradients. In these conditions, the Boltzmann equations \eqref{2.7} may be solved by the CE method \cite{CC70} conveniently adapted to account for the inelasticity in collisions. As said before, this method assumes the existence of a normal solution such that all space and time dependence of the distribution functions $f_i(\mathbf{r}, \mathbf{v},t)$ only occurs through the hydrodynamic fields. On the other hand, as noted in previous papers on granular mixtures \cite{GD02}, there
is more flexibility in the representation of the heat and mass fluxes for multicomponent systems. Even
in the case of elastic collisions, several different (but equivalent)
choices of hydrodynamic fields are used and so, some care is required in
comparing transport coefficients in the different representations. As in the undriven case \cite{GD02}, here we take the concentration $x_{1}$, the hydrostatic pressure $p=nT$, the temperature $T$, and the $d$ components of the local flow velocity $\mathbf{U}$ as the $d+3$ independent fields of the two-component mixture. Consequently, for times longer than the mean free time, the distributions $f_i(\mathbf{r}, \mathbf{v},t)$  adopt the normal form
\beq
\label{4.0}
f_i(\mathbf{r}, \mathbf{v},t)=f_i\left[\mathbf{v}|x_1(\mathbf{r},t), p(\mathbf{r},t), T(\mathbf{r},t), \mathbf{U}(\mathbf{r},t)\right].
\eeq
The notation on the right hand side indicates a functional dependence on concentration, pressure, temperature and flow velocity. In the case of small spatial variations, the functional dependence \eqref{4.0} can be made local
in space and time through an expansion in gradients of the fields. To
generate the expansion, $f_{i}$ is written as a series expansion in a formal
parameter $\epsilon $ measuring the nonuniformity of the system, i.e.,
\begin{equation}
f_{i}=f_{i}^{(0)}+\epsilon \,f_{i}^{(1)}+\epsilon^2 \,f_{i}^{(2)}+\cdots \;,
\label{4.0.1}
\end{equation}
where each factor of $\epsilon $ means an implicit gradient of a
hydrodynamic field. Moreover, in ordering the different level of approximations in the kinetic equations, one has to characterize the magnitude of the driven parameters $\gamma_\text{b}$ and $\xi_\text{b}^2$ relative to the gradients as well. As in our study \cite{GCV13} for monocomponent gases, given that both driven parameters do not induce any flux in the system, they are taken to be of zeroth order in the gradients. A different consideration must be given to the term proportional to the velocity difference $\Delta \mathbf{U}$ in Eq.\ \eqref{2.7} since it is expected that this term contributes to the mass flux in sedimentation problems, for instance. In fact, the term $\Delta \mathbf{U}$ can be interpreted as an external field (like gravity) and so, it should be considered at least to be of first order in perturbation expansion.

The time derivatives of the fields are also expanded as $\partial_{t}=\partial _{t}^{(0)}+\epsilon \partial _{t}^{(1)}+\cdots $. The coefficients of the time derivative expansion are identified from the balance equations (\ref{2.18})--(\ref{2.20}) with a representation of the fluxes and the cooling rate in the macroscopic balance equations as a similar series through their definitions as functionals of the distributions $f_{i}$. This is the usual CE method for solving kinetic equations.

\subsection{Zeroth-order distribution function}

To zeroth order in $\epsilon$, the kinetic equation \eqref{2.7} for $f_i^{(0)}$ becomes
\begin{equation}
\partial_{t}^{(0)}f_i^{(0)} -\frac{\gamma_\text{b}}{m_i^\beta}
\frac{\partial}{\partial\mathbf{v}}\cdot \mathbf{V}
f_i^{(0)}-\frac{1}{2}\frac{\xi_\text{b}^2}{m_i^\lambda}\frac{\partial^2}{\partial v^2}f_i^{(0)}=\sum_{j=1}^2 J_{ij}[f_i^{(0)},f_j^{(0)}]. \label{4.1}
\end{equation}
The balance equations at this order give
\begin{equation}
\label{4.2}
\partial_t^{(0)}x_1=0, \quad \partial_t^{(0)}\mathbf{U}=\textbf{0},
\end{equation}
\begin{equation}
\label{4.3}
T^{-1}\partial_t^{(0)}T=p^{-1}\partial_t^{(0)}p=-\Lambda^{(0)},
\end{equation}
where
\beq
\label{4.4}
\Lambda^{(0)}  = 2\gamma_\text{b}\sum_{i=1}^2
\frac{x_i \chi_i}{m_i^\beta}-\frac{\xi_\text{b}^2}{p}\sum_{i=1}^2\frac{\rho_i}{m_i^\lambda}+\zeta^{(0)}.
\eeq
Here, the cooling rate $\zeta ^{(0)}$ is determined by Eq.\ (\ref{2.17}) to zeroth order. In the Maxwellian approximation \eqref{3.19} to $\varphi_{i}$, $\zeta_i^{(0)}$ is given by Eqs. \eqref{3.14} and \eqref{3.20} with the replacements $x_{1,\text{s}}\to x_1(\mathbf{r},t)$, $p_{\text{s}}\to p(\mathbf{r},t)$ and $T_{\text{s}}\to T(\mathbf{r},t)$. In Eqs.\ \eqref{4.2} and \eqref{4.3} use has been made of the isotropic property of $f_i^{(0)}$ which leads to $\mathbf{j}_i^{(0)}=\mathbf{q}^{(0)}=\mathbf{0}$ and $P_{\alpha\beta}^{(0)}=p\delta_{\alpha\beta}$.

Since $f_i^{(0)}$ is a normal solution, then the time derivative in Eq.\ \eqref{4.1} can be represented more usefully as
\beq
\label{4.5}
\partial_t^{(0)}f_i^{(0)}=-\Lambda^{(0)}(T\partial_T+p\partial_p)f_i^{(0)}.
\eeq
Substitution of Eq.\ \eqref{4.5} into Eq.\ \eqref{4.1} yields
\begin{eqnarray}
& & -\Lambda^{(0)}(T\partial_T+p\partial_p)f_i^{(0)}
-\frac{\gamma_\text{b}}{m_i^\beta}
\frac{\partial}{\partial\mathbf{ v}}\cdot \mathbf{V}
f_i^{(0)}
\nonumber\\
&-&\frac{1}{2}\frac{\xi_\text{b}^2}{m_1^\lambda}\frac{\partial^2}{\partial v^2}f_i^{(0)}=\sum_{j=1}^2\;J_{ij}[f_i^{(0)},f_j^{(0)}]. \label{4.7}
\end{eqnarray}

The steady solution to Eq.\ \eqref{4.7} corresponds to $\Lambda^{(0)}=0$ and has been previously analyzed in Sec.\ \ref{sec3}. On the other hand, as noted in the driven monocomponent case \cite{GCV13}, for given values of $\gamma_\text{b}$, $\xi_\text{b}^2$ and $\alpha_{ij}$, the steady state condition ($\Lambda^{(0)}=0$) establishes a mapping between the partial densities, the pressure, and the temperature. Since the densities $n_i(\mathbf{r},t)$, the pressure $p(\mathbf{r},t)$, and the granular temperature  $T(\mathbf{r},t)$ are specified separately in the \emph{local} reference states $f_i^{(0)}$, the collisional cooling $\zeta^{(0)}$ is only partially compensated for by the heat injected in the system by the driving force. Thus, the time derivatives $\partial_t^{(0)}T$ and $\partial_t^{(0)}p$ are in general both \emph{different} from zero and so, the zeroth-order distribution functions $f_i^{(0)}$ depend on time through its dependence on $p$ and $T$. However, for the sake of simplicity, one could impose the steady-state condition at any point of the system, i.e.,  $\partial_t^{(0)}p=\partial_t^{(0)}T=0$. This was the choice proposed in previous theoretical works \cite{G09} in the case of the stochastic thermostat ($\gamma_\text{b}=0$ and $\lambda=0$) where the relation $(\rho/p)\xi_\text{b}^2=\zeta^{(0)}$ was assumed to hold \emph{locally}. The fact that both $\partial_t^{(0)}p\neq 0$ and $\partial_t^{(0)}T \neq 0$ gives rise to conceptual and practical difficulties not present in the previous works \cite{G09}. As we will show later, while the expression of the shear viscosity coefficient is the same in both choices ($\partial_t^{(0)}\neq 0$ and $\partial_t^{(0)}=0$), the forms of the transport coefficients associated to the mass and heat fluxes are clearly different in both choices.

In the \emph{unsteady} state, the zeroth-order distribution function $f_i^{(0)}$ obeys Eq.\ \eqref{4.1}. Dimensional analysis requires that $f_i^{(0)}$ is also given by the scaled form \eqref{3.9}, except that here the thermal velocity $v_0$ and the (reduced) model parameters $\gamma^*$ and $\xi^*$ are defined as in Sec.\ \ref{sec3} (see Eqs.\ \eqref{3.10} and \eqref{3.10.1}) with the replacements $n_\text{s}\to p(\mathbf{r},t)/T(\mathbf{r},t)$ and $T_\text{s}\to T(\mathbf{r},t)$. Thus, the zeroth-order distribution $f_i^{(0)}$ can be written as
\begin{equation}
\label{4.8}
f_{i}^{(0)}(\mathbf{r},\mathbf{ v},t)=x_i(\mathbf{r},t) \frac{p(\mathbf{r},t)}{T(\mathbf{r},t)}v_0(\mathbf{r},t)^{-d}
\varphi_{i}\left(x_1,\mathbf{c}, \gamma^*, \xi^* \right),
\end{equation}
where now $\mathbf{c}= \mathbf{V}/v_0$. The dependence of $f_i^{(0)}$ on the temperature $T$ and the pressure $p$ is not only explicit but also through $\mathbf{c}$, $\gamma^*$, and $\xi^*$. Thus,
\begin{equation}
\label{4.11}
T\partial_T f_i^{(0)}=-f_i^{(0)}-\frac{1}{2}\frac{\partial}{\partial\mathbf{v}}\cdot \left(\mathbf{V}f_i^{(0)}\right)-
\frac{1}{2}\xi^*\frac{\partial f_i^{(0)}}{\partial \xi^*}+\frac{2}{3} \omega^*\frac{\partial f_i^{(0)}}{\partial \omega^*},
\end{equation}
\begin{equation}
\label{4.12}
p\partial_p f_i^{(0)}=f_i^{(0)}-\xi^*\frac{\partial f_i^{(0)}}{\partial \xi^*}-\frac{2}{3} \omega^*\frac{\partial f_i^{(0)}}{\partial \omega^*}.
\end{equation}
Upon deriving Eqs.\ \eqref{4.11} and \eqref{4.12} use has been made of the relation $\gamma^*=\omega^*\xi^{*1/3}$, where $\omega^*$ is defined by the second identity in Eq.\ \eqref{4.10} with the change $n_\text{s}\to p(\mathbf{r},t)/T(\mathbf{r},t)$. According to Eqs.\ \eqref{4.11} and \eqref{4.12}, one has
\beq
\label{4.13}
\left(T\partial_T+p\partial_p \right) f_i^{(0)}=-\frac{1}{2}\frac{\partial}{\partial\mathbf{v}}\cdot \left(\mathbf{V}f_i^{(0)}\right)-
\frac{3}{2}\xi^*\frac{\partial f_i^{(0)}}{\partial \xi^*}.
\eeq

In dimensionless form, Eq.\ \eqref{4.1} finally becomes
\beqa
\label{4.14}
& & \Lambda^* \left(
\frac{1}{2}\frac{\partial}{\partial\mathbf{ c}}\cdot \mathbf{ c}\varphi_i+\frac{3}{2}\xi^*\frac{\partial \varphi_1}{\partial \xi^*}\right)-\frac{\omega^* \xi^{*1/3}}{M_i^\beta}\frac{\partial}{\partial \mathbf{ c}}\cdot \mathbf{ c} \varphi_i\nonumber\\
& &
- \frac{1}{4}\frac{\xi^*}{M_i^\lambda} \frac{\partial^2}{\partial c^2}\varphi_i=\sum_{i=1}^2
J_{ij}^*[\varphi_i,\varphi_j],
\eeqa
where $J_{ij}^*$ is defined by Eq.\ \eqref{3.15} and
\beq
\label{4.15}
\Lambda^{*}= \frac{\Lambda^{(0)}}{\nu_0}, \quad \nu_0=\frac{ p \sigma_{12}^{d-1}v_0}{T}.
\eeq
The partial temperature ratios $\chi_i$ can be obtained by multiplying both sides of Eq.\ \eqref{4.14} by $c^2$ and integrating over velocity. The result is
\beq
\label{4.16}
\frac{3}{2} \Lambda^* \xi^* \frac{\partial \chi_i}{\partial \xi^*}=\chi_i \Lambda^*-\Lambda_i^*,
\end{equation}
where $\Lambda^*=x_1\Lambda_1^*+x_2 \Lambda_2^*$ and
\begin{equation}
\label{4.16}
\Lambda^{*}_i = 2 \omega^* \xi^{*1/3} \frac{\chi_i}{M_i^\beta}
- \frac{\xi^*}{M_i^{\lambda-1}} +\chi_i \zeta_{i,0}^*.
\end{equation}
Here, $\zeta_{i,0}^*=\zeta_{i}^{(0)}/\nu_0$ is defined by Eq.\ \eqref{3.16} with the replacements $\chi_{i,\text{s}}\to \chi_i$ and $\varphi_{i,\text{s}}\to \varphi_i$. Approximate forms for the partial cooling rates $\zeta_{i,0}^*$ are given by Eq.\ \eqref{3.20}. The zeroth-order contribution $\zeta_0^*=\zeta^{(0)}/\nu_0$ to the cooling rate is $\zeta_0^*=x_1\chi_1 \zeta_{1,0}^*+x_2\chi_2 \zeta_{2,0}^*$.

In the steady-state ($\Lambda^*_i=0$), Eqs.\ \eqref{4.16} for $i=1,2$ agree with Eqs.\ \eqref{3.17}. In general, Eqs.\ \eqref{4.16} must be solved numerically to get the dependence of the temperature ratios $\chi_i$ on $x_1$, $\gamma^*$ and $\xi^*$. As we will show below, the transport coefficients of the mixture depend on the derivatives $\partial \chi_i/\partial x_1$, $\partial \chi_i/\partial \omega^*$ and $\partial \chi_i/\partial \xi^*$. Analytical expressions of these derivatives in the steady state limit have been obtained in Appendix \ref{appA}.

\section{Transport coefficients}
\label{sec5}

The analysis to first order in spatial gradients is more involved and follows similar steps as those worked out before for driven monodisperse gases \cite{GCV13} and undriven granular mixtures \cite{GD02}. Some technical details on the determination of the transport coefficients are provided in Appendices \ref{appB} and \ref{appC}. The form of the first-order velocity distribution functions $f_i^{(1)}$ are given by
\beqa
f_{i}^{(1)}&=&{\boldsymbol{\cal A}}_{i}\cdot \nabla x_{1}+{\boldsymbol{\cal B}}_{i}\cdot
\nabla p+{\boldsymbol{\cal C}}_{i}\cdot \nabla T\nonumber\\
& & +{\cal D}_{i,k\ell}\frac{1}{2}\left( \nabla_{k}U_{\ell}+\nabla_{\ell}U_{k}
-\frac{2}{d}\delta_{k\ell}\nabla \cdot\mathbf{U} \right)\nonumber\\
& & +{\cal E}_i \nabla\cdot \mathbf{U}+{\boldsymbol{\cal G}}_{i}\cdot \Delta \mathbf{U},  \label{5.1}
\eeqa
where the quantities ${\boldsymbol{\cal A}}_{i}(\mathbf{V})$, ${\boldsymbol{\cal B}}_{i}(\mathbf{V})$, ${\boldsymbol{\cal C}}_{i}(\mathbf{V})$, ${\cal D}_{i,k\ell}(\mathbf{V})$, ${\cal E}_i(\mathbf{V})$, and ${\boldsymbol{\cal G}}_{i}(\mathbf{V})$ are the solutions of the linear integral equations \eqref{b19}--\eqref{b24}, respectively.

However, as pointed out in the monocomponent case \cite{GCV13}, the evaluation of the transport coefficients from the above integral equations requires to know the complete time dependence of the first order corrections to the mass, momentum and heat fluxes. This is quite an intricate problem. On the other hand, some simplifications occur if attention is payed to linear deviations from the steady state described in Sec.\ \ref{sec2}. Thus, since the irreversible fluxes are already of first order in the deviations from the steady state, then one only needs to evaluate the transport coefficients to zeroth order in the deviations, namely, when the steady-state condition $\Lambda^{(0)}=0$ applies. In this case, the set of coupled linear integral equations \eqref{b19}--\eqref{b24} becomes, respectively
\beqa
\label{5.2}
& & -\frac{\gamma_\text{b}}{m_1^\beta} \frac{\partial}{\partial\mathbf{v}}\cdot \mathbf{V}
{\boldsymbol{\cal A}}_{1}-\frac{1}{2}\frac{\xi_\text{b}^2}{m_1^\lambda}\frac{\partial^2}{\partial v^2}{\boldsymbol{\cal A}}_{1}
+{\cal L}_{1} {\boldsymbol{\cal A}}_{1}+{\cal M}_{1}{\boldsymbol{\cal A}}_{2} \nonumber \\
&&+\left[\xi_b^2\frac{1}{T} \frac{m_2^{\lambda-1}-m_1^{\lambda-1}}{(m_1m_2)^{\lambda-1}}-2\gamma_\text{b} \frac{m_2^\beta-m_1^\beta}{(m_1m_2)^\beta}\left(\chi_1+x_1\frac{\partial \chi_1}{\partial x_1} \right)
\right.\nonumber\\
& & \left. -\frac{\partial \zeta^{(0)}}{\partial x_1} \right]
\left(p{\boldsymbol{\cal B}}_{1}+T{\boldsymbol{\cal C}}_{1}\right)=\mathbf{A}_1 ,
\eeqa
\beqa
\label{5.3}
& & -\frac{\gamma_\text{b}}{m_1^\beta} \frac{\partial}{\partial\mathbf{V}}\cdot \mathbf{V} {\boldsymbol{\cal B}}_{1}-\frac{1}{2}\frac{\xi_\text{b}^2}{m_1^\lambda}\frac{\partial^2}{\partial V^2}{\boldsymbol{\cal B}}_{1} +{\cal L}_{1} {\boldsymbol{\cal B}}_{1}+{\cal M}_{1}{\boldsymbol{\cal B}}_{2} \nonumber \\
& & -\left(
2\gamma_\text{b}\sum_{i=1}^2 \frac{x_i\chi_i}{m_i^\beta}+2\gamma_\text{b} p\frac{m_2^\beta-m_1^\beta}{(m_1m_2)^\beta}x_1 \frac{\partial \chi_1}{\partial p} \right.\nonumber\\
 & & \left.-\xi_b^2\frac{1}{T}\sum_{i=1}^2\frac{x_i}{m_i^{\lambda-1}}+\zeta^{(0)}+p\frac{\partial \zeta^{(0)}}{\partial p}\right){\boldsymbol{\cal B}}_{1}=\mathbf{B}_1\nonumber\\
& & +\left(2\gamma_\text{b} T\frac{m_2^\beta-m_1^\beta}{(m_1m_2)^\beta}x_1 \frac{\partial \chi_1}{\partial p}+T\frac{\partial \zeta^{(0)}}{\partial p}\right){\boldsymbol{\cal C}}_{1},
\eeqa
\beqa
\label{5.4}
& & -\frac{\gamma_\text{b}}{m_1^\beta} \frac{\partial}{\partial\mathbf{V}}\cdot \mathbf{V} {\boldsymbol{\cal C}}_{1}-\frac{1}{2}\frac{\xi_\text{b}^2}{m_1^\lambda}\frac{\partial^2}{\partial V^2}{\boldsymbol{\cal C}}_{1} +{\cal L}_{1} {\boldsymbol{\cal C}}_{1}+{\cal M}_{1}{\boldsymbol{\cal C}}_{2} \nonumber \\
& & -\left(2\gamma_\text{b} \sum_{i=1}^{2}\frac{x_i\chi_i}{m_i^\beta}+2\gamma_\text{b} T\frac{m_2^\beta-m_1^\beta}{(m_1m_2)^\beta}x_1 \frac{\partial \chi_1}{\partial T}\right.\nonumber\\
& & \left.+\zeta^{(0)}+T\frac{\partial \zeta^{(0)}}{\partial T}\right){\boldsymbol{\cal C}}_{1}=\mathbf{C}_1+\left(2\gamma_\text{b} p\frac{m_2^\beta-m_1^\beta}{(m_1m_2)^\beta}x_1 \frac{\partial \chi_1}{\partial T}\right.\nonumber\\
& & \left.+\xi_b^2\frac{p}{T^2}\sum_{i=1}^2\frac{x_i}{m_i^{\lambda-1}}+p\frac{\partial \zeta^{(0)}}{\partial T}\right){\boldsymbol{\cal B}}_{1},
\eeqa
\beqa
\label{5.5}
& & -\frac{\gamma_\text{b}}{m_1^\beta}
\frac{\partial}{\partial\mathbf{v}}\cdot \mathbf{V}
{\cal D}_{1,k\ell}-\frac{1}{2}\frac{\xi_\text{b}^2}{m_1^\lambda}\frac{\partial^2}{\partial v^2}
{\cal D}_{1,k\ell}
+{\cal L}_{1} {\cal D}_{1,k\ell}\nonumber\\
& & +{\cal M}_{1}{\cal D}_{2,k\ell}=D_{1,k\ell},
\eeqa
\beq
\label{5.6}
-\frac{\gamma_\text{b}}{m_1^\beta}
\frac{\partial}{\partial\mathbf{v}}\cdot \mathbf{V}
{\cal E}_{1}-\frac{1}{2}\frac{\xi_\text{b}^2}{m_1^\lambda}\frac{\partial^2}{\partial v^2}
{\cal E}_{1}
+{\cal L}_{1} {\cal E}_{1}+{\cal M}_{1}{\cal E}_{2}=E_{1},
\eeq
\beq
 \label{5.7}
-\frac{\gamma_\text{b}}{m_1^\beta} \frac{\partial}{\partial\mathbf{V}}\cdot \mathbf{V} {\boldsymbol{\cal G}}_{1}-\frac{1}{2}\frac{\xi_\text{b}^2}{m_1^\lambda}\frac{\partial^2}{\partial V^2}{\boldsymbol{\cal G}}_{1} +{\cal L}_{1} {\boldsymbol{\cal G}}_{1}+{\cal M}_{1}{\boldsymbol{\cal G}}_{2} =\textbf{G}_1.
\eeq
The coefficients $\mathbf{A}_1$, $\mathbf{B}_1$, $\mathbf{C}_1$, ${\cal D}_{1,k\ell}$, $\mathbf{E}_1$ and $\mathbf{G}_1$ are functions of the peculiar velocity $\mathbf{V}$ and the hydrodynamic fields. Their explicit forms are given by Eqs.\ \eqref{b10}--\eqref{b15}, respectively. Moreover, the linear operators ${\cal L}_1$ and ${\cal M}_{1}$ are defined as \begin{equation}
{\cal L}_{1}X=-\left(
J_{11}[f_{1}^{(0)},X]+J_{11}[X,f_{1}^{(0)}]+
J_{12}[X,f_{2}^{(0)}]\right) \;,
\label{5.8}
\end{equation}
\begin{equation}
{\cal M}_{1}X=-J_{12}[f_{1}^{(0)},X].  \label{5.9}
\end{equation}
The corresponding integral equations for ${\boldsymbol{\cal A}}_2$, ${\boldsymbol{\cal B}}_2$, ${\boldsymbol{\cal C}}_2$, ${\cal D}_{2,\alpha\beta}$, ${\cal E}_2$, and ${\boldsymbol{\cal G}}_2$ can be easily inferred from Eqs.\ \eqref{5.2}--\eqref{5.7} by setting $1\leftrightarrow 2$. In Eqs.\ \eqref{5.2}--\eqref{5.7}, it is understood that all the quantities are evaluated in the steady state.

Use of Eq.\ \eqref{5.1} in the definitions \eqref{2.21}--\eqref{2.23} of the fluxes gives the following forms for them to first order in gradients:
\begin{equation}
\mathbf{ j}_{1}^{(1)}=-\left(\frac{m_{1}m_{2}n}{\rho }\right) D\nabla x_{1}-\frac{
\rho}{p}D_{p}\nabla p-\frac{\rho}{T}D_T\nabla T-D_U \Delta \mathbf{U}, \label{5.10}
\end{equation}
\begin{equation}
\mathbf{q}^{(1)}=-T^{2}D^{\prime \prime}\nabla x_{1}-L\nabla p-\kappa \nabla T-\kappa_U \Delta \mathbf{U},
\label{5.11}
\end{equation}
\begin{equation}
P_{\alpha \beta}^{(1)}=-\eta \left( \partial_{\beta
}U_{\alpha}+\partial_{\alpha}U_{\beta}-\frac{2}{d}\delta_{\alpha \beta}
\nabla \cdot \mathbf{U}\right). \label{5.12}
\end{equation}
The transport coefficients in Eqs.\ \eqref{5.10}--\eqref{5.12} are
\begin{equation}
\left(
\begin{array}{c}
D \\
D_p \\
D_{T} \\
D_U \\
D''\\
L\\
\kappa  \\
\kappa_U \\
\eta
\end{array}
\right) =\left(
\begin{array}{c}
\text{diffusion coefficient} \\
\text{pressure diffusion coefficient} \\
\text{thermal diffusion coefficient} \\
\text{velocity diffusion coefficient} \\
\text{Dufour coefficient} \\
\text{pressure energy coefficient} \\
\text{thermal conductivity} \\
\text{velocity conductivity} \\
\text{shear viscosity}
\end{array}
\right)   \label{5.13}
\end{equation}
The transport coefficients associated with the mass flux $\mathbf{ j}_{1}^{(1)}$ are identified as
\begin{equation}
D=-\frac{\rho }{d m_{2}n}\int d\mathbf{ v}\,\mathbf{V} \cdot {\boldsymbol{\cal A}}_{1},  \label{5.14}
\end{equation}
\begin{equation}
D_{p}=-\frac{m_{1}p}{d\rho}\int d\mathbf{ v}\,\mathbf{V}\cdot {\boldsymbol{\cal B}}_{1},  \label{5.15}
\end{equation}
\begin{equation}
D_T=-\frac{m_{1}T}{d\rho }\int d\mathbf{ v}\;\mathbf{V}\,\cdot {\boldsymbol{\cal C}}_{1}.  \label{5.16}
\end{equation}
\begin{equation}
D_U=-\frac{m_1}{d}\int d\mathbf{v}\;\mathbf{V}\,\cdot {\boldsymbol{\cal G}}_{1}.  \label{5.17}
\end{equation}
The transport coefficients for the heat flux $\mathbf{q}^{(1)}$ are
\begin{equation}
D^{\prime \prime }=-\frac{1}{dT^{2}}\sum_{i=1}^2\,\int d\mathbf{ v}\,\frac{1}{2}
m_{i}V^{2}\mathbf{V}\cdot {\boldsymbol{\cal A}}_{i},  \label{5.18}
\end{equation}
\begin{equation}
L=-\frac{1}{d}\sum_{i=1}^2\,\int d\mathbf{ v}\,\frac{1}{2}m_{i}V^{2}\mathbf{V}
\cdot {\boldsymbol{\cal B}}_{i},  \label{5.19}
\end{equation}
\begin{equation}
\kappa =-\frac{1}{d}\sum_{i=1}^2\,\int d\mathbf{ v}\,\frac{1}{2}m_{i}V^{2}
\mathbf{V}\cdot {\boldsymbol{\cal C}}_{i}.  \label{5.20}
\end{equation}
\begin{equation}
\kappa_U =-\frac{1}{d}\sum_{i=1}^2\,\int d\mathbf{ v}\,\frac{1}{2}m_{i}V^{2}
\mathbf{V}\cdot {\boldsymbol{\cal G}}_{i}.  \label{5.21}
\end{equation}
Finally, the shear viscosity is
\begin{equation}
\eta =-\frac{1}{(d-1)(d+2)}\sum_{i=1}^2\,\int d\mathbf{ v}\, m_{i}V_{k}V_{\ell}
{\cal D}_{i,k\ell}.  \label{5.22}
\end{equation}

The evaluation of the complete set of transport coefficients is a quite long task. Here, we will focus on the transport coefficients associated to the mass flux and the shear viscosity coefficient. To determine them, we will consider the leading terms in a Sonine polynomial expansion to the unknowns ${\boldsymbol{\cal A}}_{i}$, ${\boldsymbol{\cal B}}_{i}$, ${\boldsymbol{\cal C}}_{i}$, ${\cal D}_{i,\alpha\beta}$, ${\cal E}_i$, and ${\boldsymbol{\cal G}}_{i}$. The procedure is described in Appendix \ref{appC} and only the final expressions will be provided here.

\subsection{Diffusion transport coefficients}

In dimensionless form, the diffusion transport coefficients $D$, $D_p$ and $D_T$ can be written as
\begin{equation}
D=\frac{\rho T}{m_{1}m_{2}\nu_{0}}D^*,\quad
D_{p}=\frac{nT}{\rho\nu_{0}}D_{p}^{\ast},\quad
D_T=\frac{nT}{\rho \nu_{0}}D_T^*,  \label{5.23}
\end{equation}
where $\nu_0$ is the effective frequency defined in Eq.\ \eqref{4.15}. The explicit forms are
\begin{equation}
\label{5.24}
D_p^*=\frac{a_{23}a_{30}-a_{33}a_{20}}{a_{23}a_{32}-a_{22}a_{33}},
\end{equation}
\begin{equation}
\label{5.25}
D_T^*=\frac{a_{32}a_{20}-a_{22}a_{30}}{a_{23}a_{32}-a_{22}a_{33}},
\end{equation}
\begin{equation}
\label{5.26}
D^*=\frac{a_{10}-a_{12}(D_p^*+D_T^*)}{a_{11}},
\end{equation}
where the coefficients $a_{ij}$ are defined by Eqs.\ \eqref{c8}--\eqref{c16}. The velocity diffusion coefficient $D_\text{U}$ is simply given by
\begin{equation}
\label{5.27}
D_\text{U}=\frac{\rho_1\rho_2}{\rho}\frac{\omega^*\xi^{*1/3}}{a_{11}}
\overline{m}^{\beta}\frac{m_2^\beta-m_1^\beta}{(m_1 m_2)^\beta}.
\end{equation}
Since ${\bf j}_{1}^{(1)}=-{\bf j}_{2}^{(1)}$ and $\nabla
x_{1}=-\nabla x_{2}$, $D$ must be symmetric while $D_{p}$, $D_T$, and $D_\text{U}$ must be antisymmetric with respect to the exchange $1\leftrightarrow 2$ . This can be easily verified by
noting that $x_{1}\chi_{1}+x_{2}\chi_{2}=1$.

\subsection{Shear viscosity coefficient}

The shear viscosity coefficient $\eta$ can be written as
\begin{equation}
\label{5.28}
\eta=\frac{p}{\nu_0}\left(x_1\chi_1^2\eta_1^*+x_2\chi_2^2\eta_2^*\right),
\end{equation}
where the expression of the (dimensionless) partial contributions
$\eta_i^*$ ($i=1,2$) is
\begin{equation}
\label{5.29}
\eta_1^*=\frac{\chi_1^{-1}(\tau_{22}+2\mu_{12}^{\beta}\omega^*\xi^{*1/3})-\chi_2^{-1}\tau_{12}}
{
(\tau_{11}+2\mu_{21}^{\beta}\omega^*\xi^{*1/3})(\tau_{22}+2\mu_{12}^{\beta}\omega^*\xi^{*1/3})
-\tau_{12}\tau_{21}}.
\end{equation}
The partial shear viscosity $\eta_2^*$ can be easily obtained by just making
the changes $1 \leftrightarrow 2$. The expressions of the (reduced) collision frequencies $\tau_{ij}$ are given by Eqs.\eqref{c24}--\eqref{c25}.

\section{Some illustrative driven systems}
\label{sec6}

The results derived in Sec.\ \ref{sec5} for the diffusion transport coefficients and the shear viscosity depend on the driven parameters $\gamma_\text{b}$ and $\xi_\text{b}^2$, the concentration $x_1$, and the mechanical parameters of the mixture (masses, sizes and coefficients of restitution). Moreover, they also depend on the parameters $\beta$ and $\lambda$ characterizing the class of model considered. An exploration of the full parameter space is straightforward but beyond the scope of this presentation. In this section we will consider some specific situations where a careful analysis of the impact of the parameters of the system on transport can be easily assessed.

\subsection{Tracer limit}

We consider first the special case in which one of the components of the mixture (say, for instance, species $1$) is present in tracer concentration ($x_1\to 0$). In this situation, an inspection of the coefficients $a_{ij}$ defining the diffusion coefficients shows that both $a_{20}$ and $a_{30}$ go to zero and consequently, the pressure diffusion $D_p$ and thermal diffusion $D_T$ coefficients tend to zero. The only nonzero coefficient is the (reduced) tracer diffusion coefficient $D^*$ given by
\begin{equation}
\label{6.1}
D^*=\frac{\chi_1}{\nu_D+\mu_{21}^{\beta}\omega^*\xi^{*1/3}},
\end{equation}
where in the tracer limit $\nu_D$ (defined in Eq.\ \eqref{c17}) is
\begin{equation}
\label{6.2}
\nu_D\to \frac{2\pi^{(d-1)/2}}{d\Gamma\left(\frac{d}{2}\right)}
(1+\alpha_{12})\mu_{21}\sqrt{\mu_{12}+\mu_{21}\chi_1}.
\end{equation}

Equations \eqref{6.1} and \eqref{6.2} apply for arbitrary values of the mass ratio $m_1/m_2$. In the Brownian limit ($m_1/m_2 \to \infty$), Sarracino \emph{et. al} \cite{SVCP10} have derived an expression for the self-diffusion coefficient $\overline{D}$ defined as
\beq
\label{6.3}
\overline{D}=\frac{T_2 D^*}{m_1\nu_0}.
\eeq
An explicit form for $\overline{D}$ can be easily obtained after taking the limit $m_1/m_2 \to \infty$ in our Eq.\ \eqref{6.1} for $D^*$. The result is
\begin{equation}
\label{6.4}
\overline{D}=\frac{m_1^{\beta-1}T_1}{\gamma_\text{g}+m_2^{\beta-1}\gamma_\text{b}},
\end{equation}
where $\gamma_\text{g}$ is defined in Eq.\ \eqref{3.22}. When $\lambda=2$, $\beta=1$ and for hard disks ($d=2$), Eq.\ \eqref{6.4} is the same as the one obtained from the Langevin equation.

\begin{figure}
\includegraphics[width=.45\textwidth]{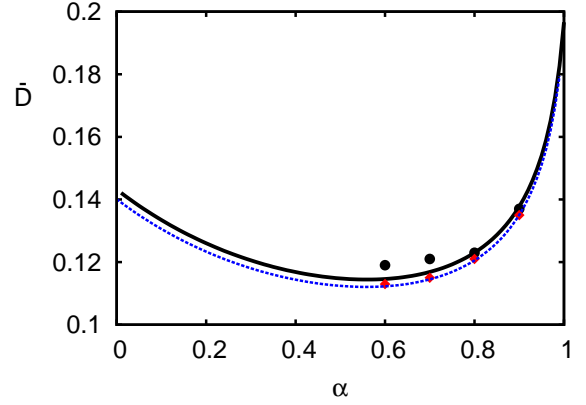}
\caption{(Color online) Plot of the self-diffusion coefficient $\overline{D}$ as a function of the (common) coefficient of restitution $\alpha\equiv \alpha_{22}=\alpha_{12}$ for a two-dimensional system ($d=2$). The parameters of the system are $m_1=100m_2$, $\sigma_1=\sigma_2$, $\phi=0.00785$, $\xi_b^2=0.2$, and $\gamma_\text{b}=0.1$. Symbols are the simulation results obtained in Ref. \cite{SVCP10} by means of DSMC method (red diamonds) and MD simulations (black circles).
The solid line is the theoretical result obtained from Eq.\ \eqref{6.1} for $m_1=100m_2$ while the dashed line corresponds to the theoretical result obtained from Eq. \eqref{6.4} in the Brownian limit ($m_1/m_2\to \infty$).}
\label{fig2}
\end{figure}

In the tracer limit, the shear viscosity of the mixture coincides with that of the excess component. Thus, when $x_1\to 0$, $\chi_2 \to 1$, and Eqs.\ \eqref{5.28}--\eqref{5.29} reduce to
\beq
\label{6.5}
\eta=\frac{p}{\nu_\eta+\frac{2\gamma_\text{b}}{m_2^\beta}},
\eeq
where
\beq
\label{6.6}
\nu_\eta=\frac{\sqrt{2}\pi^{(d-1)/2}}{d(d+2)\Gamma\left(\frac{d}{2}\right)}(3+2d-3\alpha_{22})(1+\al_{22})
n_2\sigma_2^{d-1}\sqrt{\frac{2T_2}{m_2}}.
\eeq
Equations \eqref{6.5} and \eqref{6.6} agree with the results obtained by Hayakawa \cite{H03} in the Fokker-Planck model ($\beta=1$) for monocomponent granular gases. This shows the consistency of our results with those previously derived.

\begin{figure}[!h]
  \includegraphics[width=.45\textwidth]{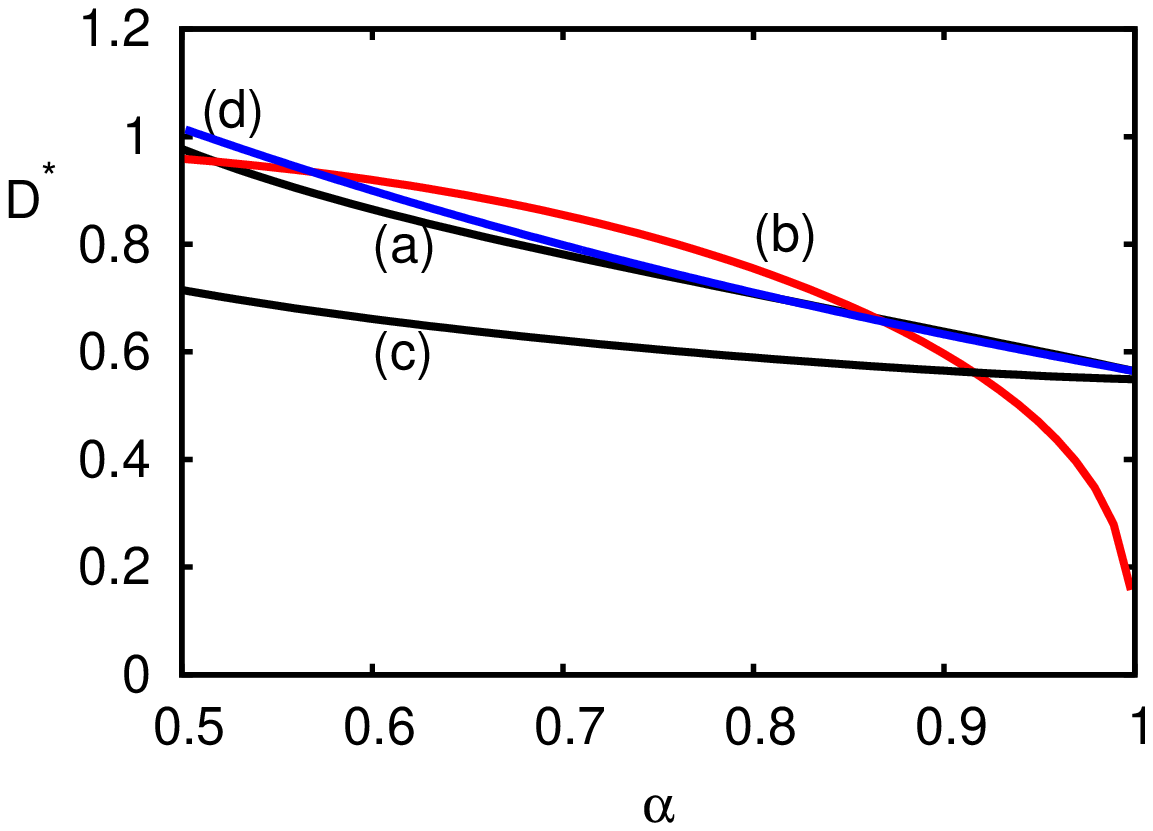}
  \includegraphics[width=.45\textwidth]{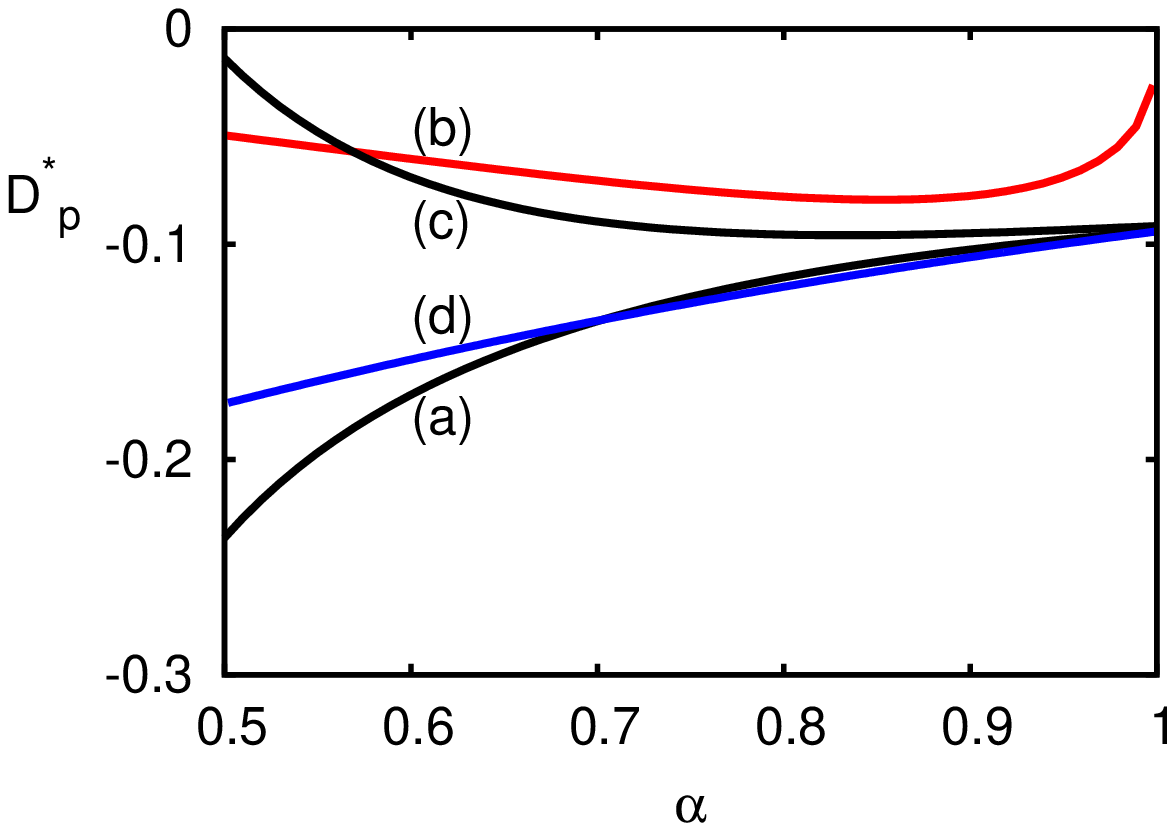}
  \includegraphics[width=.45\textwidth]{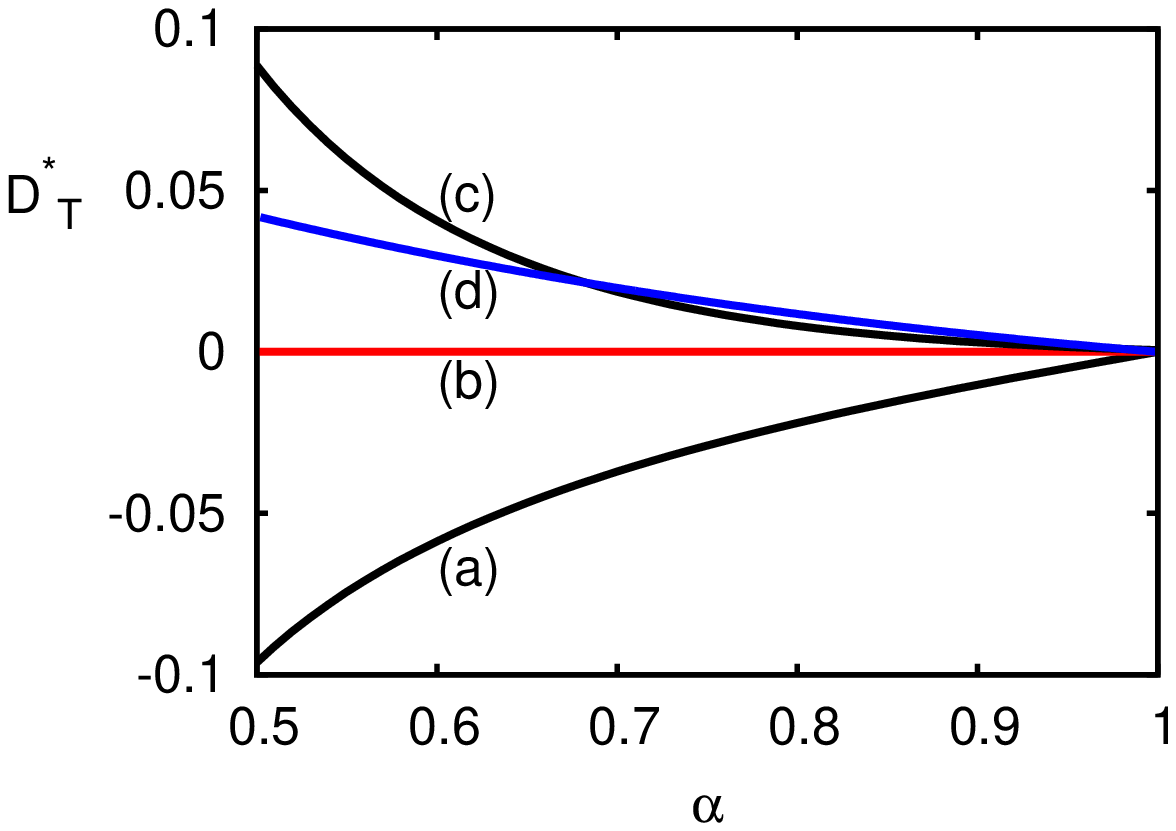}
\caption{(Color online) Reduced diffusion coefficients $D^*, \ D^*_p$, and $D^*_T$ as a function of the (common) coefficient of restitution $\alpha_{11}=\alpha_{12}=\alpha_{22}\equiv \alpha$ for an equimolar binary mixture ($x_1=\frac{1}{2}$) of hard disks ($d=2$) with   $\sigma_{1}/\sigma_{2}=1$ and  $m_1/m_2=2$. Different driven systems are plotted: (a) global stochastic thermostat ($\gamma_\text{b}=0,\ \lambda=0$), (b) local stochastic thermostat ($\gamma_\text{b}=0,\ \lambda=0$), (c) stochastic bath with friction ($\xi_\text{b}^2=0.2,\ \gamma_\text{b}=0.1, \lambda=2,$ and $\beta=1$) and (d) undriven system ($\xi_\text{b}^2=\gamma_\text{b}=0$).}
\label{fig3}
\end{figure}

The self-diffusion coefficient $\overline{D}$ is plotted in Fig.\ \ref{fig2} as a function of the (common) coefficient of restitution $\alpha$ for $d=2$. The solid line is the theoretical prediction following from Eq.\ \eqref{6.1} while the dashed line is the theoretical result obtained from Eq.\ \eqref{6.4} (Brownian limit). Symbols are DSMC results and MD simulations carried out in Ref.\ \cite{SVCP10}. There is an excellent agreement between DSMC and Brownian theory, while MD simulations present a small discrepancy with the latter at small values of the coefficient of restitution. This discrepancy is in part mitigated by the results obtained from the Boltzmann-Lorentz description (Eq.\ \eqref{6.1}), specially for strong dissipation (say for instance, $\alpha \lesssim 0.7$). Moreover, in contrast to the free cooling case \cite{GD99}, we observe that the diffusion coefficient $\overline{D}$ shows a non-monotonic behavior with a minimum at low values of $\alpha$. The lack of simulation data at small values of $\alpha$ prevent us to make a comparison in this range of inelasticity.

\subsection{Stochastic thermostat}

We consider now a system only driven by the stochastic term of thermostat (namely, when $\gamma_\text{b}\to 0$ but keeping $\gamma_\text{b}T_\text{b}$ finite). This driven system has been widely studied in the literature \cite{ernst}, specially for homogeneous monocomponent granular gases. Moreover, expressions for the diffusion transport coefficients of a granular binary mixture have been also obtained \cite{G09} for this sort of thermostat (with $\lambda=0$) when the steady-state condition $\Lambda^{(0)}=0$ applies at any point of the system (local stochastic thermostat). These expressions are displayed in Appendix \ref{appD} for the sake of completeness.

In this case ($\gamma_\text{b}=0, \; \lambda=0$), the steady-state condition simply reduces to
\begin{equation}
\label{6.7}
\xi^*=\frac{\zeta_0^*}{\sum_{i=1}^2 x_i M_i}=\frac{n\overline{m}}{\rho}\zeta_0^*,
\end{equation}
while the temperature ratio is determined from the condition
\begin{equation}
\label{6.8}
m_2\chi_1 \zeta_{1,0}^*=m_1 \chi_2 \zeta_{2,0}^*.
\end{equation}
Thus, according to Eq.\ \eqref{6.7}, the noise strength $\xi^*$ is a function of the coefficients of restitution and the parameters of the mixture. The diffusion transport coefficients are
\beq
\label{6.9}
D_p^*=\frac{\frac{x_1}{2}\frac{n\overline{m}}{\rho}\zeta_0^{*2}\frac{\partial \chi_1}{\partial \xi^*}\delta-(\nu_D-\frac{\zeta_0^*}{2}\delta)
(x_1\chi_1-\frac{\rho_1}{\rho}-x_1\zeta_0^*\frac{\partial \chi_1}{\partial \xi^*})}
{\frac{1}{2}\zeta_0^{*2}\delta^2-(\nu_D-\zeta_0^*\delta)(\nu_D-\frac{1}{2}\zeta_0^*\delta)},
\eeq
\beq
\label{6.10}
D_T^*=\frac{\frac{x_1}{2}\xi^{*}\frac{\partial \chi_1}{\partial \xi^*}(\nu_D-\zeta_0^*\delta)-\frac{\zeta_0^*}{2}\delta (x_1\chi_1-\frac{\rho_1}{\rho}-x_1\zeta_0^*\frac{\partial \chi_1}{\partial \xi^*})}{\frac{1}{2}\zeta_0^{*2}\delta^2-(\nu_D-\zeta_0^*\delta)(\nu_D-\frac{1}{2}\zeta_0^*\delta)},
\eeq
\beq
\label{6.11}
D^*=\frac{\chi_1+x_1\frac{\partial \chi_1}{\partial x_1}-\left[\frac{(m_1-m_2)n}{\rho}\zeta_0^*-\frac{\partial \zeta_0^*}{\partial x_1}\right](D_p^*+D_T^*)}{\nu_D},
\eeq
where $\delta = 1-\frac{n\overline{m}}{\rho}(\partial \zeta_0^*/\partial \xi^*)$ and $\nu_D$ is given by Eq.\ \eqref{c17}. In addition, since $D_U\propto \gamma^*=\omega^*\xi^{*1/3}$ (see Eq.\ \eqref{5.27}), the velocity diffusion coefficient $D_U$ vanishes in the case of the stochastic thermostat.
\begin{figure}[!h]
  \includegraphics[width=.45\textwidth]{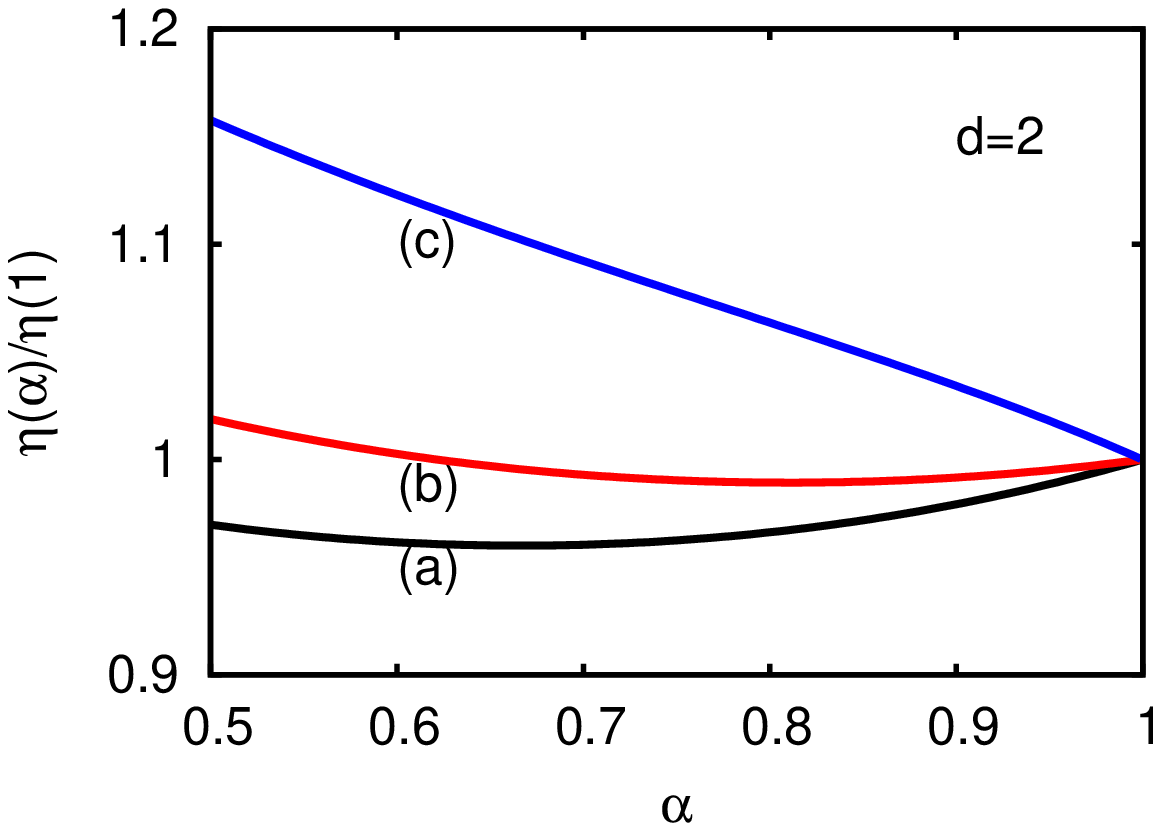}
  \includegraphics[width=.45\textwidth]{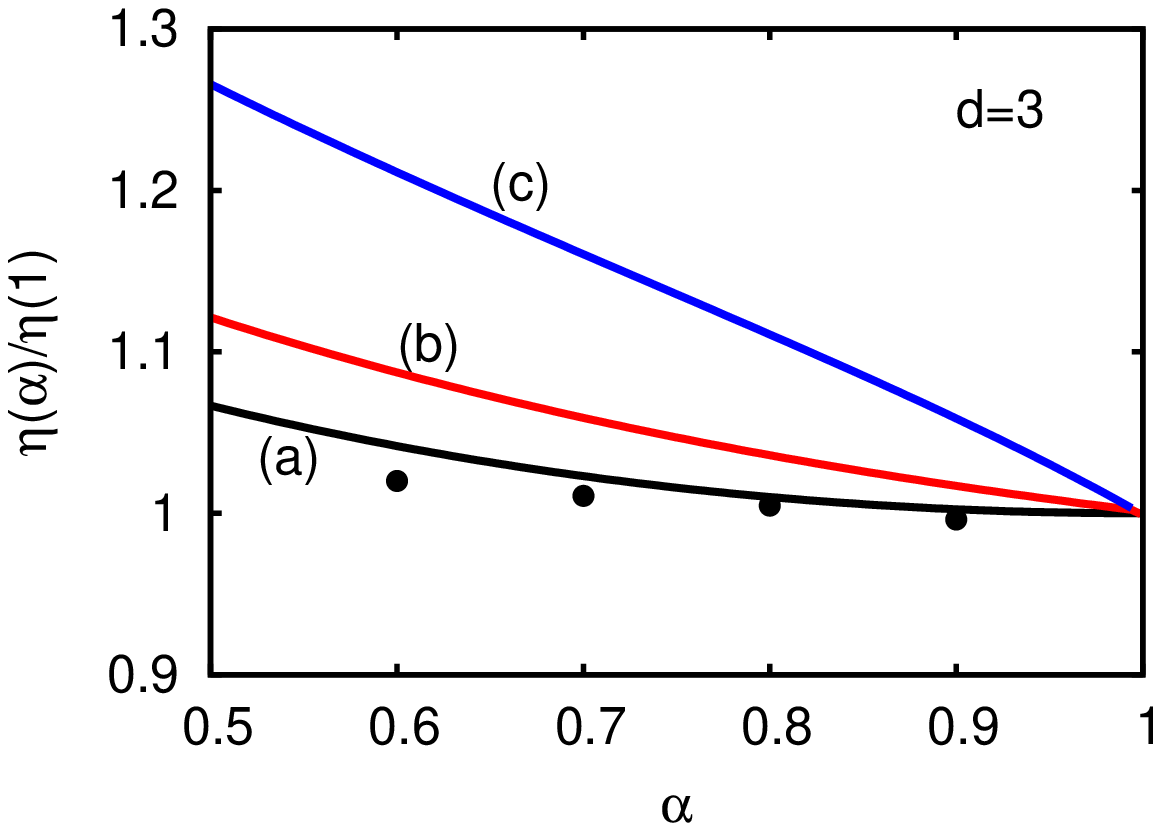}
\caption{(Color online) Plot of the (reduced) shear viscosity coefficient $\eta(\alpha)/\rojo\eta(1)$ versus the (common) coefficient of restitution $\alpha_{11}=\alpha_{12}=\alpha_{22}\equiv \alpha$ for an equimolar binary mixture ($x_1=\frac{1}{2}$) of hard disks (top panel) and hard spheres (bottom panel) with $\sigma_{1}/\sigma_{2}=1$ and three different values of the mass ratio: (a) $m_1/m_2=1$, (b) $m_1/m_2=2$, and (c) $m_1/m_2=4$. The lines correspond to the theoretical results derived for the stochastic thermostat ($\gamma_\text{b}=0,\ \lambda=0$). The symbols are the DSMC results for a mixture of mechanically equivalent particles driven by the stochastic thermostat (Ref. \cite{GM02}).}
\label{fig4}
\end{figure}
\begin{figure}
  \includegraphics[width=.45\textwidth]{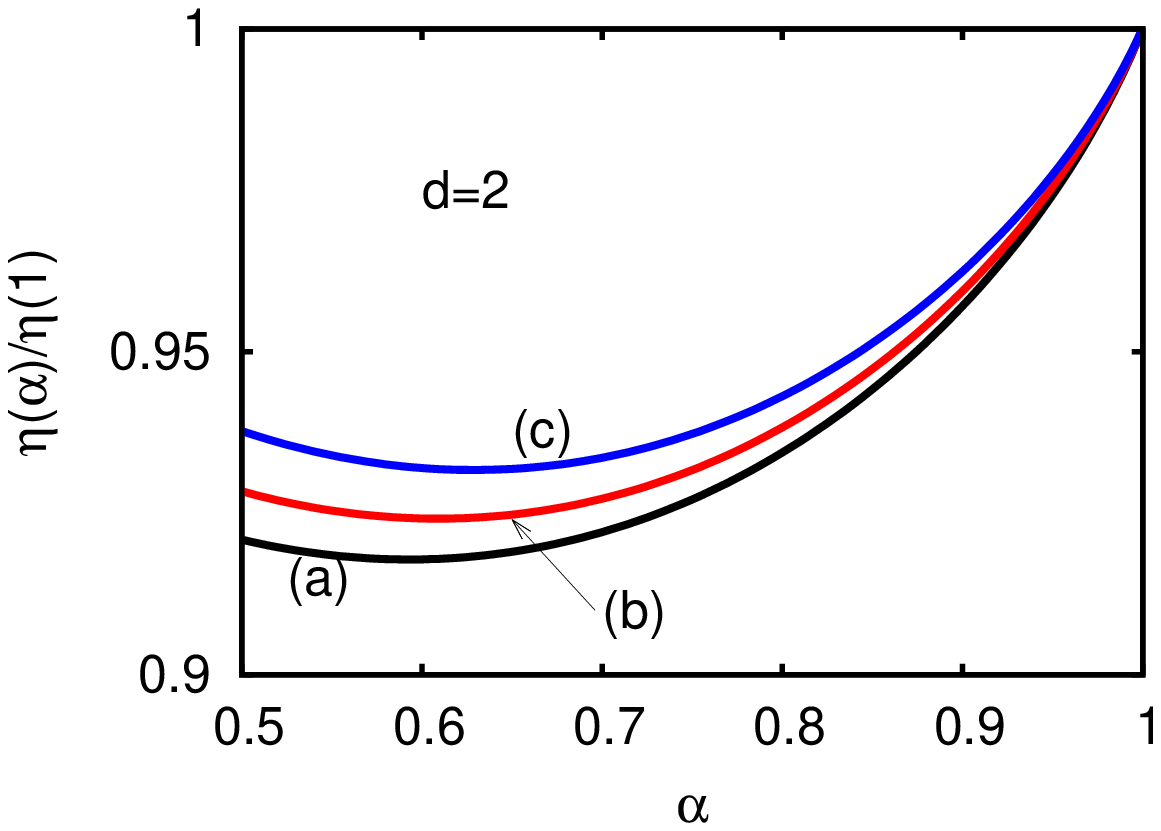}
  \includegraphics[width=.45\textwidth]{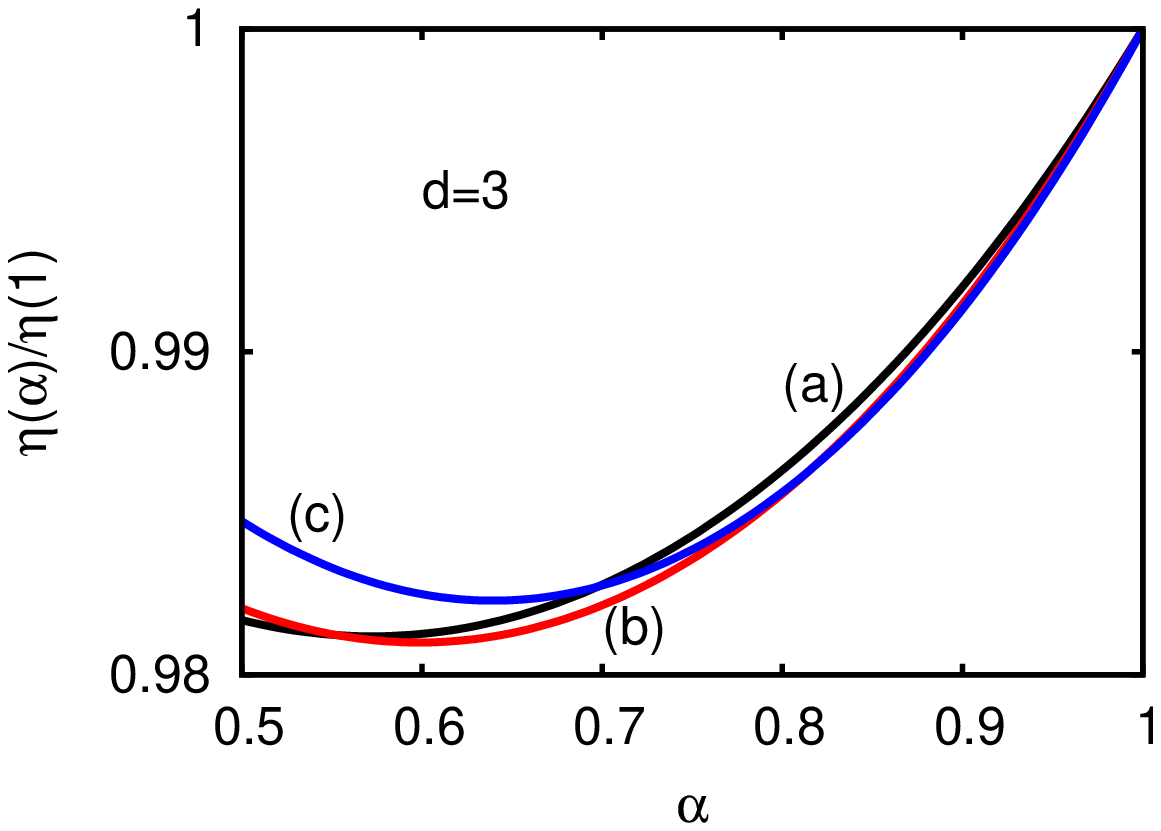}
\caption{(Color online) Plot of the (reduced) shear viscosity coefficient $\eta(\alpha)\eta/\eta(1)$ versus the (common) coefficient of restitution $\alpha_{11}=\alpha_{12}=\alpha_{22}\equiv \alpha$ for an equimolar binary mixture ($x_1=\frac{1}{2}$) of hard disks (top panel) and hard spheres (bottom panel) with $\sigma_{1}/\sigma_{2}=1$ and three different values of the mass ratio: (a) $m_1/m_2=1$, (b) $m_1/m_2=2$, and (c) $m_1/m_2=4$. The lines are the results derived for the general driven system with model parameters $\xi_\text{b}^2=0.2,\ \gamma_\text{b}=0.1, \lambda=2$, and  $\beta=1$.}
\label{fig5}
\end{figure}

Comparison between Eqs.\ \eqref{6.9}--\eqref{6.11} with Eqs.\ \eqref{d1}--\eqref{d2} clearly shows that the forms of the diffusion coefficients obtained here differ from those previously derived \cite{G09} by using a (simple) local thermostat. In particular, while the latter choice yields a vanishing thermal diffusion coefficient $D_T$, we found here that $D_T \neq 0$. To illustrate the differences between both choices of thermostat, Fig.\ \ref{fig3} shows the (reduced) diffusion coefficients $D^*$, $D_p^*$, and $D_T^*$ as a function of the (common) coefficient of restitution $\alpha$ for an equimolar mixture ($x_1=\frac{1}{2}$) with $\sigma_1/\sigma_2=1$ and $m_1/m_2=2$. Different driven systems have been plotted. The free cooling system is also plotted for the sake of completeness. First, as expected the thermostat does not play a neutral role on mass transport since the $\alpha$-dependence of the diffusion coefficients between the driven and undriven systems is clearly different. On the other hand, at a more quantitative level, it is quite apparent that the results derived in this paper for the diffusion $D^*$ and pressure diffusion $D_p^*$ coefficients are closer to their corresponding undriven counterparts \cite{GD02} than those obtained by using the local stochastic thermostat. In fact, the theoretical predictions for both coefficients obtained from the (global) stochastic thermostat compare quite well with the free cooling results even for quite strong values of dissipation (say for instance, $\alpha \gtrsim 0.7$). The biggest discrepancy between both theories is for the thermal diffusion coefficient $D_T^*$ since while this transport coefficient is negative in the driven case, it becomes positive in the undriven case. The change of sign of $D_T^*$ could have some implications in processes related to thermal diffusion segregation \cite{BKD13,G13}.

The shear viscosity coefficient $\eta$ is given by Eq.\ \eqref{5.28} where the partial contributions $\eta_i^*$ are
\beq
\label{6.12}
\eta_1^*=\frac{\chi_1^{-1}\tau_{22}-\chi_2^{-1}\tau_{12}}{\tau_{11}\tau_{22}-\tau_{12}\tau_{21}}, \quad
\eta_2^*=\frac{\chi_2^{-1}\tau_{11}-\chi_1^{-1}\tau_{21}}{\tau_{11}\tau_{22}-\tau_{12}\tau_{21}}.
\eeq
Although the expression of $\eta$ for a driven granular mixture has not been previously derived, a simple inspection of the integral equation \eqref{5.5} shows that the form \eqref{6.12} also holds for the case of the local stochastic thermostat. Figure \ref{fig4} shows the dependence of the ratio $\eta(\alpha)/\eta(1)$ on $\alpha$ for $\sigma_1/\sigma_2=1$, $x_1=\frac{1}{2}$ and for several values of the mass ratio ($m_1/m_2=1, 2$ and 4). Here, $\eta(1)$ is the shear viscosity of the binary mixture for elastic collisions. Some DSMC data obtained in Ref.\ \cite{GM02} for a single granular gas ($m_1=m_2$) of inelastic hard spheres have been also included. A good agreement with theory is observed. We also see that the deviation of $\eta$ from its functional form for elastic collisions is less significant than the one found for undriven mixtures \cite{MG03,GM07}. Moreover, except for a single gas of hard disks, we observe that the shear viscosity of a driven granular mixture increases with respect to its elastic value as the inelasticity increases.

\subsection{General driven system: stochastic bath with friction}

The analysis of the general case ($\xi_\text{b}^2\neq 0$ and $\gamma_\text{b} \neq 0$) is more difficult than when the system is only driven by the stochastic thermostat. This is specially apparent for the diffusion coefficients $D^*$, $D_p^*$ and $D_T^*$ since their evaluation requires to get all the derivatives of the temperature ratio (i.e., the derivatives of $\chi_1$ with respect to $\xi^*$, $\omega^*$, and $x_1$) in the vicinity of the steady state. To illustrate the behavior of the diffusion coefficients and the shear viscosity, we have considered an equimolar binary mixture driven by the model parameters $\xi_\text{b}^2=0.2$ and $\gamma_\text{b}=0.1$ with $\lambda=2$ and  $\beta=1$.

The $\alpha$-dependence of the (reduced) diffusion coefficients for the above driven system has been also included in Fig.\ \ref{fig3}. We observe that the behavior of these coefficients is in general quite different to that of the stochastic thermostat, specially in the cases of the pressure diffusion $D_p^*$ and the thermal diffusion $D_T^*$ coefficients. Thus, while both coefficients increase as $\alpha$ decreases in the general case ($\xi_\text{b}^2\neq 0$ and $\gamma_\text{b} \neq 0$), the opposite happens for the stochastic thermostat ($\xi_\text{b}^2\neq 0$ but $\gamma_\text{b}=0$). On the other hand, the dependence of the diffusion coefficient $D^*$ on the coefficient of restitution is qualitatively similar in both driven systems since $D^*$ increases with increasing inelasticity.

Finally, we analyze in Fig.\ \ref{fig5} the shear viscosity of the mixture. As in Fig.\ \ref{fig4}, we plot  $\eta(\alpha)/\eta(1)$ as a function of the (common) coefficient of restitution. We observe that the influence of dissipation on $\eta$ for the general case is opposite to the one found in Fig.\ \ref{fig4} for the stochastic thermostat since the ratio $\eta(\alpha)/\alpha(1)$ decreases with decreasing $\alpha$ in the former case. Thus, the main effect of inelasticity of collisions when the granular mixture is fluidized by the combination of a stochastic bath with friction is to inhibit its momentum transport with respect to the elastic collision case. However, the deviation of $\eta$ from its elastic value is much smaller than the one obtained for the diffusion coefficients since  the inelastic shear viscosity differs less than 2\% from its corresponding elastic form $\eta(1)$.

\section{Discussion}
\label{sec7}

The main objective of this work has been to determine the transport coefficients of a granular binary mixture driven by a stochastic bath with friction. The results have been obtained from the set of nonlinear (inelastic) Boltzmann equations for the mixture and are expected to apply at low densities. The derivation of the hydrodynamic equations consists of two steps. First, the macroscopic balance equations \eqref{2.18}--\eqref{2.20} for the partial densities, the total momentum, and energy are obtained from the set of coupled Boltzmann equations \eqref{2.7}. Then, the fluxes and the cooling rate appearing in these hydrodynamic equations have been determined from a solution of the Boltzmann equations by means of the CE method. Their forms have been expressed in terms of the hydrodynamic fields and their spatial gradients. The corresponding constitutive equations for the mass, heat, and momentum fluxes to first order in spatial gradients are given by Eqs.\ \eqref{5.10}--\eqref{5.12}, respectively, and the associated transport coefficients are defined by Eqs.\ \eqref{5.14}--\eqref{5.17} for the mass flux, Eqs.\ \eqref{5.18}--\eqref{5.21} for the heat flux, and   Eq.\ \eqref{5.22} for the pressure tensor. It is worthwhile noticing that all the above results are \emph{exact} within the framework of the Boltzmann equation.

As in the undriven case \cite{GD02}, the transport coefficients are given in terms of the solution of the set of coupled linear integral equations \eqref{5.2}--\eqref{5.7}. A practical evaluation of these coefficients requires the truncation of a Sonine polynomial expansion. Thus, although these results are approximated, they are not limited in principle to weak inelasticity and apply to arbitrary values of the coefficients of restitution, the mass and size ratios, and the concentration of the mixture. In addition, they also depend on the driven parameters $\gamma_\text{b}$ (which represents the friction coefficient of the drag force) and $\xi_\text{b}^2$ (which represents the strength of the stochastic force). The explicit determination of the complete set of transport coefficients (nine coefficients) as functions of the full parameter space is beyond the scope of this paper and we have focused here on the diffusion and the shear viscosity coefficients.

\begin{figure}
  \includegraphics[width=.45\textwidth]{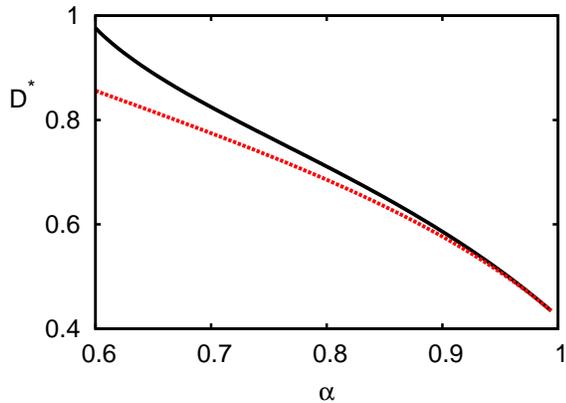}
\caption{(Color online) Plot of the (reduced) diffusion coefficient $D^*$ as a function of the (common) coefficient of restitution
$\alpha_{11}=\alpha_{12}=\alpha_{22}\equiv \alpha$ for an equimolar binary mixture ($x_1=\frac{1}{2}$) of hard disks with $\sigma_{1}/\sigma_{2}=2$ and $m_1/m_2=8$. Here, the mixture is driven by a global stochastic thermostat ($\gamma_\text{b}=0,\ \lambda=0$). The solid line is the result derived from Eq.\ \eqref{6.11} while the dashed line has been obtained by neglecting the derivatives of $\chi_1$ and $\zeta_0^*$ with respect to $\xi^*$ in Eq.\ \eqref{6.11}.   }
\label{fig6}
\end{figure}

As pointed out in our previous effort \cite{GCV13} for monocomponent gases, a subtle point is the generalization of the driving external forces (which are usually introduced in homogeneous situations) to \emph{inhomogeneous} states. This is a quite important issue since one has to consider first small perturbations to steady homogeneous states to determine the fluxes from the CE solution and then, identify the corresponding transport coefficients. These quantities are intrinsic properties of the driven granular mixture. Although the above generalization is a matter of choice, it has important implications on the form of the transport coefficients \cite{GMT13}. For the sake of simplicity, in previous works carried out by one of the authors of the present paper \cite{G09}, it was assumed that the external driving force has the same expression as in the homogeneous case, except that the parameters of the force are chosen to get stationary values of the pressure $p$ and temperature $T$ of the mixture in the CE zeroth-order approximation (i.e., $\partial_t^{(0)}p=\partial_t^{(0)}T=0$). Nevertheless, this is a particular choice for the perturbations since in general it is expected that the pressure and temperature are specified separately in the local reference state $f_i^{(0)}$ of each species and so, $p$ and $T$ are in general time-dependent quantities (i.e., $\partial_t^{(0)}p\neq 0$ and $\partial_t^{(0)}T\neq 0$). This latter feature gives rise to new technical difficulties in the evaluation of the transport coefficients since one would need in particular to numerically integrate the differential equations verifying some velocity moments of the distributions $f_i^{(0)}$ to get the time dependence of the transport coefficients. This is quite an intricate problem. On the other hand, since we are interested here in evaluating the fluxes in the first order in the deviations from the steady homogeneous state, the transport coefficients associated to the mass, momentum and heat fluxes can be determined to zeroth-order in the deviations (steady-state conditions). As said before, in this paper we have explicitly obtained the transport coefficients associated to the mass flux and the pressure tensor. Their explicit forms are given by Eqs.\ \eqref{5.24}--\eqref{5.27} for the diffusion coefficients $D$, $D_p$, $D_T$, and $D_U$, respectively, and Eqs.\ \eqref{5.28}--\eqref{5.29} for the shear viscosity coefficient $\eta$.

The expressions derived for the set $\left\{D, D_p, D_T, D_U, \eta\right\}$ clearly show the complex dependence of these coefficients on the concentration, the mechanical parameters of the mixture (masses, diameters and coefficients of restitution) and the driven model parameters $\gamma_\text{b}$ and $\xi_\text{b}^2$. Our results also indicate that while the expressions of the diffusion coefficients derived here differ from those previously obtained \cite{G09} by using a local thermostat, the form of the shear viscosity is the same for both choices of thermostat. This is an expected result since the evaluation of $\eta$ does not involve any contribution coming from the action of the operator $\partial_t^{(0)}$ on the pressure and temperature gradients. In addition, a careful evaluation of the transport coefficients for a variety of mass and diameter ratios and coefficients of restitution has shown that the impact of collisional dissipation on transport in driven mixtures is less significant than the one previously observed in undriven mixtures \cite{GD02}.

It is worthwhile to remark that, although we evaluate the transport coefficients under steady-state conditions, the time-dependence of the reference states $f_i^{(0)}$ is inherited through the derivatives of the temperature ratio $\chi_1$ and the (reduced) cooling rate $\zeta_0^*$ with respect to the (reduced) model parameters $\omega^*$ and $\xi^*$. This additional dependence can be easily seen in particular in the expressions \eqref{6.9}--\eqref{6.11} for the diffusion coefficients $D_p^*$, $D_T^*$ and $D^*$, respectively. In order to gauge the effect of those derivatives on mass transport, Fig. \ref{fig6} shows $D^*$ versus $\alpha$ as given by Eq.\ \eqref{6.11} and the result for $D^*$ by neglecting the derivatives $\partial \chi_1/\partial \xi^*$ and $\partial \zeta_0^*/\partial \xi^*$ in Eq.\ \eqref{6.11}. We have considered
a binary mixture composed by disks of the same mass density ($\sigma_1/\sigma_2=2$ and $m_1/m_2=8$) with $x_1=\frac{1}{2}$. Clearly, inclusion of those derivatives becomes more significant as the inelasticity increases.

Apart from its academic interest, we think that our results could be also relevant from a more practical point of view since many of the simulations reported \cite{puglisi, ernst} for flowing granular mixtures have considered the use of external driving forces. In this context, it is convenient to provide to simulators with the expressions of the transport coefficients when the granular mixture is driven by the sort of thermostat used here. As a matter of fact, given the lack of theoretical results covering this problem, in most of the cases the \emph{elastic} forms of the transport coefficients are used to compare simulations with theoretical results. Moreover, as pointed out in the Introduction, the driven Boltzmann equations \eqref{2.7} could be also considered as an alternative way to model bidisperse suspensions. In this context, the coefficients $\gamma_\text{b}$ and $\xi_\text{b}^2$ of the model could be adjusted to optimize the agreement with some property of interest measured in simulations or real experiments. This was the procedure followed in Ref.\ \cite{GTSH12} in the case of monodisperse gas-solid suspensions. Finally, given that the results reported in this paper are restricted to the low-density regime, the extension of the present results to dense driven systems could be an interesting project for the next future. In this case, the revised Enskog theory could be a good starting point \cite{GDH07} to determine the influence of external driven parameters on transport at moderate densities.

\acknowledgments

We thank the authors of Ref. \cite{SVCP10} for providing us their simulation results.
The present work has been supported by the Ministerio de Educaci\'on y Ciencia (Spain) through grant No. FIS2010-16587, partially financed by FEDER funds and by the Junta de Extremadura (Spain) through Grant No. GRU10158. The  research  of Nagi Khalil  has  been  supported  by  the  postdoctoral  grant FIS2008-01339.

\appendix
\section{Evaluation of the derivatives of the temperature ratio $\chi_1$ with respect to $\xi^*$, $\omega^*$, and $x_1$ in the vicinity of the steady state.}
\label{appA}

In this Appendix we will evaluate the derivatives of the temperature ratio $\chi_1$ with respect to $\xi^*$, $\omega^*$, and $x_1$ in the vicinity of the steady state. These derivatives are needed to determine the complete set of transport coefficients of the mixture. First, in order to determine $\partial \chi_1/\partial \xi^*$ we start from Eq.\ \eqref{4.16} for $i=1$:
\begin{equation}
\label{a1}
\frac{3}{2} \Lambda^* \xi^* \frac{\partial \chi_1}{\partial \xi^*}=\chi_1 \Lambda^*-\Lambda_1^*,
\end{equation}
where
\begin{equation}
\label{a2}
\Lambda^{*}_1 =2 \omega^* \xi^{*1/3} \frac{\chi_1}{M_1^\beta}
- \frac{\xi^*}{M_1^{\lambda-1}} +\chi_1 \zeta_{1,0}^*.
\end{equation}
Here, $\Lambda^*=x_1\Lambda_1^{*}+x_2 \Lambda_2^{*}$ and $\zeta_0^*=x_1 \chi_1 \zeta_{1,0}^*+
x_2 \chi_2 \zeta_{2,0}^*$. According to Eq.\ \eqref{3.20}, the dependence of $\zeta_{1,0}^*$ on $x_1$, $\omega^*$ and $\xi^*$ can be computed from the relation
\begin{equation}
\label{a3}
\zeta_{1,0}^*=\chi_1^{1/2}M_1^{-1/2}\zeta'_1(x_1,\theta),
\end{equation}
where $\theta=M_1\chi_2/(M_2\chi_1)$,  $\chi_2=(1-x_1\chi_1)/x_2$ and
\begin{eqnarray}
\label{a3a}
\zeta'_1(x_1,\theta)&=&\frac{\sqrt{2}\pi^{(d-1)/2}}{d\Gamma\left(\frac{d}{2}\right)}
x_1\left(\frac{\sigma_1}{\sigma_{12}}\right)^{d-1} (1-\alpha_{11}^2 ) \nn
  &+&\frac{4\pi^{(d-1)/2}}{d\Gamma\left(\frac{d}{2}\right)} x_2 \mu_{21}\left(1+\theta \right)^{1/2} (1+\alpha_{12}) \nn  &\times & \left[1-\frac{1}{2}\mu_{21}(1+\alpha_{12})(1+\theta)\right].
\end{eqnarray}

At the steady state, $\Lambda_1^*=\Lambda_2^*=\Lambda^*=0$, and so one has to take care in Eq.\ \eqref{a1} since the expression of
the derivative $\partial \chi_1^*/\partial \xi^*$ becomes indeterminate. This difficulty can be fixed by means of l'Hopital's rule. In this case, we take first the derivative with respect to $\xi^*$ in both sides of Eq.\ \eqref{a1} and then take the steady-state limit. The result is
\begin{equation}
\label{a4}
\frac{\partial \chi_1}{\partial \xi^*}=\frac{\frac{\partial \Lambda^*}{\partial \xi^*}\chi_1-
\frac{\partial \Lambda_1^*}{\partial \xi^*}}{\frac{3}{2}\xi^*\frac{\partial \Lambda^*}{\partial \xi^*}},
\end{equation}
where it is understood that all the derivatives are evaluated at the steady state. The derivatives appearing in the numerator and denominator of Eq.\ \eqref{a4} can be expressed in terms of the unknown $\Delta = (\partial \chi_1^*/\partial \xi^*)_\text{s}$. Here, the subindex $\text{s}$ means that the derivative is evaluated in the steady state. After some algebra, it is straightforward to see that $\Delta$ obeys the quadratic equation
\begin{eqnarray}
\label{a5}
\frac{3}{2}\xi^*\Lambda_{1}^{(\xi)} \Delta^2&+&\left(\frac{3}{2}\xi^*\Lambda_{0}^{(\xi)}-
\chi_1\Lambda_{1}^{(\xi)}+\Lambda_{11}^{(\xi)}\right)\Delta\nonumber\\
& & +\Lambda_{10}^{(\xi)}-\chi_1\Lambda_{0}^{(\xi)}=0,
\end{eqnarray}
where $\Lambda_{0}^{(\xi)}=x_1\Lambda_{10}^{(\xi)}+x_2\Lambda_{20}^{(\xi)}$, $\Lambda_{1}^{(\xi)}=x_1\Lambda_{11}^{(\xi)}+x_2\Lambda_{21}^{(\xi)}$, and
\beq
\label{a6}
\Lambda_{10}^{(\xi)}=\frac{2}{3}\omega^* \xi^{*-2/3} \frac{\chi_1}{M_1^\beta}-M_1^{1-\lambda},
\eeq
\beq
\label{a7}
\Lambda_{11}^{(\xi)}=\frac{2 \omega^* \xi^{*1/3}}{M_1^\beta}+\frac{3}{2}\zeta_{1,0}^*-\chi_1^{-1/2}\frac{M_1^{1/2}}{x_2M_2}
\frac{\partial \zeta_1'}{\partial \theta},
\eeq
\beq
\label{a8}
\Lambda_{20}^{(\xi)}=\frac{2}{3}\omega^* \xi^{*-2/3} \frac{\chi_2}{M_2^\beta}-M_2^{1-\lambda},
\eeq
\beq
\label{a9}
\Lambda_{21}^{(\xi)}=-\frac{x_1}{x_2}\frac{2 \omega^* \xi^{*1/3}}{M_2^\beta}-\frac{3}{2}\frac{x_1}{x_2}\zeta_{2,0}^*
-\frac{M_1}{x_2M_2^{3/2}}\frac{\chi_2^{3/2}}{\chi_1^2}\frac{\partial \zeta_2'}{\partial \theta}.
\eeq
An analysis of the solutions to Eq.\ \eqref{a5} shows that in general one of the roots leads to un-physical behavior of the diffusion coefficients in the quasielastic limit. We take the other root as the the physical root of the quadratic equation \eqref{a5}.

Once the derivative $\Delta$ is known, we can determine the remaining derivatives $\partial \chi_1/\partial \omega^*$ and $\partial \chi_1/\partial x_1$ in a similar way. In order to get $\partial \chi_1/\partial \omega^*$, we take first the derivative  of Eq.\ \eqref{a1} with respect to $\omega^*$ and then consider the steady-state conditions. The final result is
\begin{equation}
\label{a10}
\left(\frac{\partial \chi_1}{\partial \omega^*}\right)_\text{s}=\frac{\chi_1\Lambda_{0}^{(\gamma)}-\Lambda_{10}^{(\gamma)}-\frac{3}{2}\xi^*\Delta
\Lambda_{0}^{(\gamma)}}{\frac{3}{2}\xi^*\Delta\Lambda_{1}^{(\xi)}-\chi_1\Lambda_{1}^{(\xi)}+
\Lambda_{11}^{(\xi)}},
\end{equation}
where $\Lambda_{0}^{(\gamma)}=x_1\Lambda_{10}^{(\gamma)}+x_2\Lambda_{20}^{(\gamma)}$ and
\beq
\label{a11}
\Lambda_{10}^{(\gamma)}=2 \xi^{*1/3} \frac{\chi_1}{M_1^\beta}, \quad
\Lambda_{20}^{(\gamma)}=2 \xi^{*1/3} \frac{\chi_2}{M_2^\beta}.
\eeq
Analogously, the derivative $\partial \chi_1/\partial x_1$ is
\begin{equation}
\label{a12}
\left(\frac{\partial \chi_1}{\partial x_1}\right)_\text{s}=\frac{\chi_1\Lambda_{0}^{(x_1)}-\Lambda_{10}^{(x_1)}-\frac{3}{2}\xi^*\Delta
\Lambda_{0}^{(x_1)}}{\frac{3}{2}\xi^*\Delta\Lambda_{1}^{(\xi)}-\chi_1\Lambda_{1}^{(\xi)}+
\Lambda_{11}^{(\xi)}},
\end{equation}
where
\beq
\label{a13}
\Lambda_{10}^{(x_1)}=\chi_1^{3/2}M_1^{-1/2}\left(\frac{\partial \zeta_1'}{\partial x_1}\right)_\theta,
\eeq
\beqa
\label{a14}
\Lambda_{20}^{(x_1)}&=&\chi_2^{3/2}M_2^{-1/2}\left(\frac{\partial \zeta_2'}{\partial x_1}\right)_\theta+ \frac{\chi_2-\chi_1}{x_2}\nonumber\\
& & \times \left(2 \frac{\omega^*\xi^{*1/3}} {M_2^\beta}+\frac{3}{2}\zeta_{2,0}^*\right),
\eeqa
\beq
\label{a15}
\Lambda_{0}^{(x_1)}=\Lambda_1^*-\Lambda_2^*+x_1\Lambda_{10}^{(x_1)}+x_2\Lambda_{20}^{(x_1)}.
\eeq

\section{First order approximation}
\label{appB}

In this Appendix we provide some technical details in the derivation of the first order approximation $f_1^{(1)}$. To first order in the gradients, the equation for $f_1^{(1)}$ is
\beqa
&&\partial_{t}^{(0)}f_{1}^{(1)}-\frac{\gamma_\text{b}}{m_1^\beta}
\frac{\partial}{\partial\mathbf{ v}}\cdot \mathbf{V}
f_1^{(1)}-\frac{1}{2}\frac{\xi_\text{b}^2}{m_1^\lambda}\frac{\partial^2}{\partial v^2}f_1^{(1)}
+{\cal L}_{1} f_{1}^{(1)}\nonumber\\
& & +{\cal M}_{1}f_{2}^{(1)}
=-\left( D_{t}^{(1)}+\mathbf{V}\cdot \nabla \right)f_{1}^{(0)}+\frac{\gamma_\text{b}}{m_1^\beta} \Delta \mathbf{U}\cdot \frac{\partial}{\partial \mathbf{V}} f_1^{(0)}\;,  \nonumber\\
\label{b1}
\eeqa
where $D_{t}^{(1)}=\partial _{t}^{(1)}+\mathbf{U}\cdot \nabla $ and the linear operators ${\cal L}_{1}$ and ${\cal M}_{1}$ are defined in Eqs.\ \eqref{5.8} and \eqref{5.9}, respectively. The kinetic equation for $f_2^{(1)}$ can be easily obtained from Eq.\ \eqref{b1} by setting $1\leftrightarrow2$. The action of the operator $D _{t}^{(1)}$ on the hydrodynamic fields is
\begin{equation}
D_{t}^{(1)}x_{1}=0,  \label{b2}
\end{equation}
\begin{equation}
D_{t}^{(1)}p=-\frac{d+2}{d}p\nabla \cdot \mathbf{U}-\zeta^{(1)}p,  \label{b3}
\end{equation}
\begin{equation}
D_{t}^{(1)}T=-\frac{2}{d}T\nabla \cdot \mathbf{U}-\zeta^{(1)}T,  \label{b4}
\end{equation}
\begin{equation}
D_{t}^{(1)}\mathbf{U}=-\rho ^{-1}\nabla p-\frac{\gamma_\text{b}}{\rho} \sum_{i=1}^2\frac{
\mathbf{ j}_{i}^{(1)}}{m_{i}^\beta}-\frac{\gamma_\text{b}}{\rho}\sum_{i=1}^2\frac{\rho_i}{m_i^\beta}\Delta \mathbf{U},  \label{b5}
\end{equation}
where use has been made of the result $\mathbf{ j}_{i}^{(0)}=\mathbf{ q}^{(0)}=\mathbf{0}$. Note that in contrast to the undriven case \cite{GD02}, there is a nonzero first-order contribution $\zeta^{(1)}$ to the cooling rate. Since the cooling rate is a scalar, its corrections to first order in the gradients can arise only from the divergence of the velocity vector $\nabla \cdot \mathbf{U}$. Thus, $\zeta^{(1)}$ can be simply written as
\begin{equation}
\label{b6}
\zeta^{(1)}=\zeta_U \nabla \cdot \mathbf{U}.
\end{equation}
The time derivative $D_t^{(1)}f_{1}^{(0)}$ can be evaluated by taking into account Eqs.\ \eqref{b2}--\eqref{b5} with the result
\begin{eqnarray}
\label{b7}
& & D_{t}^{(1)} f_{1}^{(0)}
=\frac{\partial f_1^{(0)}}{\partial p} D_t^{(1)}p+
\frac{\partial f_1^{(0)}}{\partial T} D_t^{(1)}T\nonumber\\
& +& \sum_{i=1}^d\frac{\partial f_1^{(0)}}{\partial U_i} D_t^{(1)}U_i
=\rho^{-1}\frac{\partial f_1^{(0)}}{\partial \mathbf{V}}  \nabla p
\nonumber\\
& -& \left[\left(\frac{d+2}{d}+\zeta_U\right) p\frac{\partial f_1^{(0)}}{\partial p}+\left(\frac{2}{d}+\zeta_U\right) T\frac{\partial f_1^{(0)}}{\partial T}\right]\nabla\cdot \mathbf{U}  \nonumber\\
& +& \frac{\gamma_\text{b}}{\rho}\frac{\partial}{\partial \mathbf{V}} f_1^{(0)} \sum_{i=1}^2 \frac{\bf{j}_i^{(1)}}{m_i^\beta}+\frac{\gamma_\text{b}}{\rho} \sum_{i=1}^2 \frac{\rho_i}{m_i^\beta}\frac{\partial}{\partial \mathbf{V}} f_i^{(0)} \cdot \Delta \mathbf{U},\nonumber\\
\end{eqnarray}
where use has been made of the property
\begin{equation}
\label{b8}
\frac{\partial f_1^{(0)}}{\partial U_i}=-\frac{\partial f_1^{(0)}}{\partial V_i}.
\end{equation}
With the use of Eq.\ \eqref{b7}, Eq.\ \eqref{b1} can be written as
\beqa
\label{b9}
&&\partial_{t}^{(0)}f_{1}^{(1)}-\frac{\gamma_\text{b}}{m_1^\beta}
\frac{\partial}{\partial\mathbf{ v}}\cdot \mathbf{V}
f_1^{(1)}-\frac{1}{2}\frac{\xi_\text{b}^2}{m_1^\lambda}\frac{\partial^2}{\partial v^2}f_1^{(1)}
+{\cal L}_{1} f_{1}^{(1)}\nonumber\\
& & +{\cal M}_{1}f_{2}^{(1)}=\mathbf{ A}_{1}\cdot \nabla x_{1}+\mathbf{ B}_{1}\cdot \nabla p+\mathbf{ C}_{1}\cdot \nabla T\nonumber\\
& &
+D_{1,k\ell}\frac{1}{2}\left(\nabla_{k}U_{\ell}+\nabla_{\ell}U_{k}
-\frac{2}{d}\delta_{k\ell}\nabla \cdot\mathbf{U}\right)\nonumber\\
& & +E_1 \nabla \cdot \mathbf{U}+\mathbf{G}_1\cdot \Delta \mathbf{U}.
\eeqa
The coefficients of the field gradients on the right side are functions of $
\mathbf{V}$ and the hydrodynamic fields. They are given by
\begin{equation}
\mathbf{ A}_{1}(\mathbf{V})=-\mathbf{V}\frac{\partial f_1^{(0)}}{\partial x_1}+\frac{\gamma_\text{b} (m_2^\beta-m_1^\beta)}{\rho^2 (m_1 m_2)^{\beta-1}}\frac{p}{T} D \frac{\partial f_1^{(0)}}{\partial \mathbf{V}},  \label{b10}
\end{equation}
\beq
\label{b11}
\mathbf{ B}_{1}(\mathbf{V})=-\mathbf{V}\frac{\partial f_1^{(0)}}{\partial p} -\rho^{-1}
\frac{\partial f_1^{(0)}}{\partial \mathbf{V}}
+\frac{\gamma_\text{b}(m_2^\beta-m_1^\beta)}{p(m_1 m_2)^\beta} D_p \frac{\partial f_1^{(0)}}{\partial \mathbf{V}},
\eeq
\begin{equation}
\mathbf{ C}_{1}(\mathbf{V})=-\mathbf{V} \frac{\partial f_1^{(0)}}{\partial T}+\frac{\gamma_\text{b}(m_2^\beta-m_1^\beta)}{T(m_1m_2)^\beta}D_T \frac{\partial f_1^{(0)}}{\partial \mathbf{V}},  \label{b12}
\end{equation}
\begin{equation}
D_{1,k\ell}(\mathbf{V})=V_k \frac{\partial f_1^{(0)} }{\partial V_\ell},
\label{b13}
\end{equation}
\beqa
\label{b14}
E_1(\mathbf{V})&=&\left(\frac{d+2}{d}+\zeta_U\right) p \frac{\partial f_1^{(0)}}{\partial p}+\left(\frac{2}{d}+\zeta_U\right) T\frac{\partial f_1^{(0)}}{\partial T}\nonumber\\
& & +\frac{1}{d}\mathbf{V}\cdot \frac{\partial f_1^{(0)}}{\partial \mathbf{V}},
\eeqa
\begin{equation}
\label{b15}
\mathbf{G}_1(\mathbf{V})=\frac{\gamma_\text{b}}{\rho} \frac{m_2^\beta-m_1^\beta}{(m_1m_2)^\beta}\left(\rho_2+ D_U\right) \frac{\partial f_1^{(0)}}{\partial \mathbf{V}}.
\end{equation}
The solution to Eq.\ \eqref{b9} is of the form \eqref{5.1}. The coefficients
${\boldsymbol{\cal A}}_{1}$, ${\boldsymbol{\cal B}}_{1}$, ${\boldsymbol{\cal C}}_{1}$, ${\cal D}_{1,k\ell}$, ${\cal E}_1$ and ${\boldsymbol{\cal G}}_{1}$ appearing in Eq.\ \eqref{5.1} are unknown functions of the peculiar velocity. The partial temperatures and the cooling rate depend on space through their dependence on $x_1$, $p$, and $T$. The time derivative $\partial_{t}^{(0)}$ acting on ${\boldsymbol{\cal A}}_{1}$, ${\boldsymbol{\cal B}}_{1}, \ldots$  can be evaluated by the replacement $\partial_{t}^{(0)}\to -\Lambda^{(0)}(p\partial_p+T\partial_T)$. In addition, there are also contributions coming from the action of the operator $\partial _{t}^{(0)}$ on the temperature and pressure gradients. They are given by
\begin{eqnarray}
\partial_{t}^{(0)}\nabla T &=&\left[ \xi_\text{b}^2 \frac{m_2^{\lambda-1}-m_1^{\lambda-1}}{(m_1m_2)^{\lambda-1}}-T \frac{\partial \zeta^{(0)}}{\partial x_1}\right.\nonumber\\
&- & \left.2\gamma_\text{b} T\frac{m_2^\beta-m_1^\beta}{(m_1m_2)^\beta}\left(\chi_1+x_1\frac{\partial \chi_1}{\partial x_1} \right)\right] \nabla x_1 \nonumber \\
&-& \left(2\gamma_\text{b} \sum_{i=1}^{2}\frac{x_i\chi_i}{m_i^\beta}+2\gamma_\text{b} T\frac{m_2^\beta-m_1^\beta}{(m_1m_2)^\beta}x_1 \frac{\partial \chi_1}{\partial T}\right.\nonumber\\
&+&\left.\zeta^{(0)}+T\frac{\partial \zeta^{(0)}}{\partial T}\right) \nabla T \nonumber  \\
&-& \left(2\gamma_\text{b} T\frac{m_2^\beta-m_1^\beta}{(m_1m_2)^\beta}x_1 \frac{\partial \chi_1}{\partial p}+T\frac{\partial \zeta^{(0)}}{\partial p}\right) \nabla p, \nonumber\\ \label{b16}
\end{eqnarray}
\begin{eqnarray}
\partial_{t}^{(0)}\nabla p &=&\left[\frac{p}{T}\xi_\text{b}^2 \frac{m_2^{\lambda-1}-m_1^{\lambda-1}}{(m_1m_2)^{\lambda-1}}-p \frac{\partial \zeta^{(0)}}{\partial x_1}\right. \nonumber\\
&-&\left.2\gamma_\text{b} p\frac{m_2^\beta-m_1^\beta}{(m_1m_2)^\beta}\left(\chi_1+x_1\frac{\partial \chi_1}{\partial x_1} \right)\right] \nabla x_1 \nonumber \\
&-& \left(\xi_\text{b}^2\frac{p}{T^2}\sum_{i=1}^2\frac{x_i}{m_i^{\lambda-1}}+p\frac{\partial \zeta^{(0)}}{\partial T} \right. \nonumber  \\
&+&\left.2\gamma_\text{b} p\frac{m_2^\beta-m_1^\beta}{(m_1m_2)^\beta}x_1 \frac{\partial \chi_1}{\partial T}
\right) \nabla T \nonumber\\
&-& \left(2\gamma_\text{b}\sum_{i=1}^2 \frac{x_i\chi_i}{m_i^\beta}+2\gamma_\text{b} p\frac{m_2^\beta-m_1^\beta}{(m_1m_2)^\beta}x_1 \frac{\partial \chi_1}{\partial p}\right.\nonumber\\
&-&\left.\xi_\text{b}^2\frac{1}{T}\sum_{i=1}^2\frac{x_i}{m_i^{\lambda-1}} +\zeta^{(0)}+p\frac{\partial \zeta^{(0)}}{\partial p}\right) \nabla p.  \nonumber\\\label{b17}
\end{eqnarray}
Upon deriving Eqs. \eqref{b16} and \eqref{b17}, use has been made of the relations $\nabla x_1=-\nabla x_2$ and  $\nabla (x_1 \chi_1)=-\nabla (x_2 \chi_2)$.

The corresponding integral equations for the unknowns ${\boldsymbol{\cal A}}_{1}$,
${\boldsymbol{\cal B}}_{1}$, ${\boldsymbol{\cal C}}_{1}$, ${\cal D}_{1,k\ell}$, ${\cal E}_1$ and ${\boldsymbol{\cal G}}_{1}$ are identified as the coefficients of the independent gradients in Eq.\ \eqref{b9}. This yields the following set of coupled linear integral equations:
\beqa
\label{b19}
& & -\Lambda^{(0)}\left(p\frac{\partial}{\partial p}+T\frac{\partial}{\partial T}\right){\boldsymbol{\cal A}}_{1}-\frac{\gamma_\text{b}}{m_1^\beta} \frac{\partial}{\partial\mathbf{v}}\cdot \mathbf{V}
{\boldsymbol{\cal A}}_{1}-\frac{1}{2}\frac{\xi_\text{b}^2}{m_1^\lambda}\frac{\partial^2}{\partial v^2}{\boldsymbol{\cal A}}_{1}
 \nonumber \\
&&+{\cal L}_{1} {\boldsymbol{\cal A}}_{1}+{\cal M}_{1}{\boldsymbol{\cal A}}_{2}+\left[\xi_\text{b}^2\frac{1}{T} \frac{m_2^{\lambda-1}-m_1^{\lambda-1}}{(m_1m_2)^{\lambda-1}}\right.\nonumber\\
& & \left.
-2\gamma_\text{b} \frac{m_2^\beta-m_1^\beta}{(m_1m_2)^\beta}\left(\chi_1+x_1\frac{\partial \chi_1}{\partial x_1} \right)-\frac{\partial \zeta^{(0)}}{\partial x_1} \right]
\left(p{\boldsymbol{\cal B}}_{1}+T{\boldsymbol{\cal C}}_{1}\right)
\nonumber\\
& & =\mathbf{A}_1 ,
\eeqa
\beqa
\label{b20}
& & -\Lambda^{(0)}\left(p\frac{\partial}{\partial p}+T\frac{\partial}{\partial T}\right){\boldsymbol{\cal B}}_{1}-\frac{\gamma_\text{b}}{m_1^\beta} \frac{\partial}{\partial\mathbf{V}}\cdot \mathbf{V} {\boldsymbol{\cal B}}_{1}-\frac{1}{2}\frac{\xi_\text{b}^2}{m_1^\lambda}\frac{\partial^2}{\partial V^2}{\boldsymbol{\cal B}}_{1} \nonumber \\
& & +{\cal L}_{1} {\boldsymbol{\cal B}}_{1}+{\cal M}_{1}{\boldsymbol{\cal B}}_{2}-\left(2\gamma_\text{b}\sum_{i=1}^2 \frac{x_i\chi_i}{m_i^\beta}+2\gamma_\text{b} p\frac{m_2^\beta-m_1^\beta}{(m_1m_2)^\beta}x_1 \frac{\partial \chi_1}{\partial p}\right.\nonumber\\
 & & \left.-\xi_b^2\frac{1}{T}\sum_{i=1}^2\frac{x_i}{m_i^{\lambda-1}}+\zeta^{(0)}+p\frac{\partial \zeta^{(0)}}{\partial p}\right){\boldsymbol{\cal B}}_{1}=\mathbf{B}_1\nonumber\\
& & +\left(2\gamma_\text{b} T\frac{m_2^\beta-m_1^\beta}{(m_1m_2)^\beta}x_1 \frac{\partial \chi_1}{\partial p}+T\frac{\partial \zeta^{(0)}}{\partial p}\right){\boldsymbol{\cal C}}_{1},
\eeqa
\beqa
\label{b21}
& & -\Lambda^{(0)}\left(p\frac{\partial}{\partial p}+T\frac{\partial}{\partial T}\right){\boldsymbol{\cal C}}_{1}-\frac{\gamma_\text{b}}{m_1^\beta} \frac{\partial}{\partial\mathbf{V}}\cdot \mathbf{V} {\boldsymbol{\cal C}}_{1}-\frac{1}{2}\frac{\xi_\text{b}^2}{m_1^\lambda}\frac{\partial^2}{\partial V^2}{\boldsymbol{\cal C}}_{1} \nonumber \\
& &+{\cal L}_{1} {\boldsymbol{\cal C}}_{1}+{\cal M}_{1}{\boldsymbol{\cal C}}_{2} -\left(2\gamma_\text{b} \sum_{i=1}^{2}\frac{x_i\chi_i}{m_i^\beta}+2\gamma_\text{b} T\frac{m_2^\beta-m_1^\beta}{(m_1m_2)^\beta}x_1 \frac{\partial \chi_1}{\partial T}\right.\nonumber\\
& & \left.+\zeta^{(0)}+T\frac{\partial \zeta^{(0)}}{\partial T}\right){\boldsymbol{\cal C}}_{1}=\mathbf{C}_1+\left(2\gamma_\text{b} p\frac{m_2^\beta-m_1^\beta}{(m_1m_2)^\beta}x_1 \frac{\partial \chi_1}{\partial T}\right.\nonumber\\
& & \left.+\xi_b^2\frac{p}{T^2}\sum_{i=1}^2\frac{x_i}{m_i^{\lambda-1}}+p\frac{\partial \zeta^{(0)}}{\partial T}\right){\boldsymbol{\cal B}}_{1},
\eeqa
\beqa
\label{b22}
& & -\Lambda^{(0)}\left(p\frac{\partial}{\partial p}+T\frac{\partial}{\partial T}\right){\cal D}_{1,k\ell}-\frac{\gamma_\text{b}}{m_1^\beta}
\frac{\partial}{\partial\mathbf{v}}\cdot \mathbf{V}
{\cal D}_{1,k\ell}
\nonumber\\
& &-\frac{1}{2}\frac{\xi_\text{b}^2}{m_1^\lambda}\frac{\partial^2}{\partial v^2}
{\cal D}_{1,k\ell}
+{\cal L}_{1} {\cal D}_{1,k\ell} +{\cal M}_{1}{\cal D}_{2,k\ell}=D_{1,k\ell},\nonumber\\
\eeqa
\beqa
\label{b23}
& & -\Lambda^{(0)}\left(p\frac{\partial}{\partial p}+T\frac{\partial}{\partial T}\right){\cal E}_{1}-\frac{\gamma_\text{b}}{m_1^\beta}
\frac{\partial}{\partial\mathbf{v}}\cdot \mathbf{V}
{\cal E}_{1}\nonumber\\
& & -\frac{1}{2}\frac{\xi_\text{b}^2}{m_1^\lambda}\frac{\partial^2}{\partial v^2}
{\cal E}_{1}
+{\cal L}_{1} {\cal E}_{1}+{\cal M}_{1}{\cal E}_{2}=E_{1},
\eeqa
\beqa
 \label{b24}
& & -\Lambda^{(0)}\left(p\frac{\partial}{\partial p}+T\frac{\partial}{\partial T}\right){\boldsymbol{\cal G}}_{1} -\frac{\gamma_\text{b}}{m_1^\beta} \frac{\partial}{\partial\mathbf{V}}\cdot \mathbf{V} {\boldsymbol{\cal G}}_{1}\nonumber\\
& & -\frac{1}{2}\frac{\xi_\text{b}^2}{m_1^\lambda}\frac{\partial^2}{\partial V^2}{\boldsymbol{\cal G}}_{1} +{\cal L}_{1} {\boldsymbol{\cal G}}_{1}+{\cal M}_{1}{\boldsymbol{\cal G}}_{2} =\textbf{G}_1.
\eeqa

As noted in Sec.\ \ref{sec4}, in the first order of the deviations from the steady state, we only need to know the transport coefficients to zeroth order in the deviations, namely, when $\Lambda^{(0)}=0$. This implies that the first term appearing in the left-hand side of Eqs.\ \eqref{b19}--\eqref{b24} vanishes and the integral equations defining the transport coefficients are given by Eqs.\ \eqref{5.2}--\eqref{5.7}.

\section{Leading Sonine approximations}
\label{appC}

In this Appendix, we obtain the explicit expressions of the diffusion transport coefficients $D$, $D_p$, $D_T$, and $D_U$ and the shear viscosity coefficient $\eta$ in the first Sonine approximation. The diffusion coefficients are defined by Eqs.\ \eqref{5.14}--\eqref{5.17}, respectively while $\eta$ is defined by Eq.\ \eqref{5.22}. The procedure to get these coefficients is quite similar to the one previously used in the free cooling case \cite{GD02}. Only some partial results will be presented here.

In the case of the coefficients $D$, $D_p$, and $D_T$, the leading Sonine approximations
(lowest degree polynomial) of the quantities ${\boldsymbol {\cal
A}}_{i}$, ${\boldsymbol {\cal B}}_{i}$, and ${\boldsymbol {\cal
C}}_{i}$ are, respectively,
\begin{equation}
\label{c1} {\boldsymbol {\cal A}}_{1}({\bf V})\to -f_{1,M}{\bf
V}\frac{m_1m_2n D}{\rho n_1T_1},{\boldsymbol {\cal
A}}_{2}({\bf V})\to f_{2,M}{\bf V}\frac{m_1m_2n D}{\rho n_2T_2}
\end{equation}
\begin{equation}
\label{c2} {\boldsymbol {\cal B}}_{1}({\bf V})\to -f_{1,M}{\bf
V}\frac{\rho D_p}{p n_1T_1} ,{\boldsymbol {\cal B}}_{2}({\bf
V})\to f_{2,M}{\bf V}\frac{\rho D_p}{p n_2T_2},
\end{equation}
\begin{equation}
\label{c3} {\boldsymbol {\cal C}}_{1}({\bf V})\to -f_{1,M}{\bf
V}\frac{\rho D_T}{T n_1T_1}, {\boldsymbol {\cal C}}_{2}({\bf
V})\to f_{2,M}{\bf V}\frac{\rho D_T}{T n_2T_2},
\end{equation}
where $f_{i,M}$ are the Maxwellian distributions
\begin{equation}
\label{c4} f_{i,M}({\bf V})=n_i\left(\frac{m_i}{2\pi T_i}\right)^{d/2}
\exp\left(-\frac{m_i V^2}{2T_i}\right).
\end{equation}
In order to determine the above diffusion coefficients, we substitute first ${\boldsymbol {\cal
A}}_{i}$, ${\boldsymbol {\cal B}}_{i}$, and ${\boldsymbol {\cal
C}}_{i}$ by their leading Sonine approximations in Eqs.\ \eqref{5.2}--\eqref{5.4}. Then, we multiply these equations by $m_1 \mathbf{V}$ and integrates over velocity. After some algebra, the corresponding algebraic equations for the (reduced) coefficients $D^*$, $D_p^*$ and $D_T^*$ (defined by Eq.\ \eqref{5.23}) can be written as
\begin{equation}
\label{c5}
a_{11} D^* + a_{12} (D_p^*+ D_T^*)=a_{10},
\end{equation}
\begin{equation}
\label{c6}
a_{22} D_p^*+ a_{23} D_T^*=a_{20},
\end{equation}
\begin{equation}
\label{c7}
a_{32} D_p^*+ a_{33} D_T^*=a_{30},
\end{equation}
where
\begin{equation}
\label{c8}
a_{11}=\nu_D+\overline{m}^\beta\omega^* \xi^{*1/3}\frac{\rho_1 m_1^\beta+\rho_2 m_2^\beta}{\rho (m_1m_2)^\beta},
\end{equation}
\beqa
\label{c9}
a_{12}&=&-2\omega^* \xi^{*1/3}M_1^{-\beta}\frac{m_2^\beta-m_1^\beta}{m_2^\beta}\frac{
\partial }{\partial x_{1}}(x_{1}\chi_{1})\nonumber\\
&+&
\xi^*M_1^{1-\lambda}\frac{m_2^{\lambda-1}-m_1^{\lambda-1}}{m_2^{\lambda-1}}-\frac{\partial \zeta_0^*}{\partial x_1},
\eeqa
\begin{equation}
\label{c10}
a_{10}=\frac{\partial }{\partial x_{1}}(x_{1}\chi_{1}),
\end{equation}
\beq
\label{c11}
a_{23}=-2\omega^*\xi^{*1/3}M_1^{-\beta}\frac{m_2^\beta-m_1^\beta}{m_2^\beta}x_1 p\frac{\partial \chi_1}{\partial p}-\frac{p}{\nu_0}\frac{\partial \zeta^{(0)}}{\partial p},
\eeq
\beq
\label{c12}
a_{22}=a_{11}+a_{23},
\eeq
\begin{equation}
\label{c13}
a_{20}=x_1\chi_1-\frac{\rho_1}{\rho}+x_1 p \frac{\partial \chi_1}{\partial p},
\end{equation}
\beqa
\label{c14}
a_{32}&=&-2\omega^* \xi^{*1/3} M_1^{-\beta}\frac{m_2^\beta-m_1^\beta}{m_2^\beta}x_1 T\frac{\partial \chi_1}{\partial T}\nonumber\\
& & -\xi^*\sum_i x_iM_i^{1-\lambda}-\frac{T}{\nu_0}\frac{\partial \zeta^{(0)}}{\partial T},
\eeqa
\beq
\label{c15}
a_{33}=a_{11}+a_{32},
\eeq
\begin{equation}
\label{c16}
a_{30}= x_1 T \frac{\partial \chi_1}{\partial T}.
\end{equation}
In the above equations, $\nu_0$ is the effective frequency defined in the second identity of Eq.\ \eqref{4.15} and $\nu_D$ is the (reduced) collision frequency \cite{GM07}
\beqa
\label{c17}
\nu_D&=&-\frac{1}{dn_{1}T_{1}}\int d{\bf V}_{1}m_{1}{\bf V}_{1}\cdot
\left( J_{12}[{\bf v}_{1}|f_{1,M}{\bf V}_{1},f_{2}^{(0)}]\right.\nonumber\\
& & \left.-\frac{x_1T_1}{x_2T_2}
J_{12}[{\bf v}_{1}|f_{1}^{(0)},f_{2,M}{\bf V}_{2}]\right)\nonumber\\
&=&\frac{2\pi^{(d-1)/2}}{d\Gamma\left(\frac{d}{2}\right)}
(1+\alpha_{12})
\left(\frac{\theta_1+\theta_2}{\theta_1\theta_2}\right)^{1/2}\nonumber\\
& & \times\left(x_2 M_1^{-1}+
x_1 M_2^{-1}\right).
\eeqa
The solution to Eqs.\ \eqref{c5}--\eqref{c7} is given by Eqs.\ \eqref{5.24}--\eqref{5.26}.

The coefficient $D_U$ is decoupled from the other diffusion coefficients. The leading Sonine approximations to
${\boldsymbol {\cal G}}_{1}$ and ${\boldsymbol {\cal G}}_{2}$ are
\beq
\label{c17.1}
{\boldsymbol {\cal G}}_{1}({\bf V})\to -f_{1,M}{\bf
V}\frac{D_U}{n_1T_1} ,{\boldsymbol {\cal G}}_{2}({\bf
V})\to f_{2,M}{\bf V}\frac{D_U}{n_2T_2}.
\eeq
The expression \eqref{5.27} for $D_U$ can be easily obtained from Eqs.\ \eqref{5.7} and \eqref{c17.1}.

In the case of the pressure tensor, the leading Sonine
approximation for the function ${\cal D}_{i,k\ell}$ is
\begin{equation}
\label{c18} {\cal D}_{i,k\ell}({\bf V})\to -f_{i,M}({\bf V}) \frac{\eta_{i}}{T}
R_{i,k\ell}({\bf V}),\quad i=1,2
\end{equation}
where
\begin{equation}
\label{c19}
R_{i,k\ell}({\bf V})=m_i\left( V_{k}V_{\ell}-
\frac{1}{d}V^2\delta_{k\ell}\right),
\end{equation}
and
\begin{equation}
\label{c20} \eta_i=-\frac{1}{(d-1)(d+2)}\frac{T}{n_iT_i^2}\int
d{\bf v} R_{i,k\ell}({\bf V}){\cal D}_{i,k\ell}({\bf V}).
\end{equation}
The shear viscosity $\eta$ is given by Eq.\ \eqref{5.28} where $\eta_i^*=\nu_0\eta_i$.
The integral equations for the
(reduced) coefficients $\eta_i^*$ are decoupled from the diffusion
transport coefficients. The two coefficients $\eta_{i}^*$ are
obtained by multiplying Eqs.\ (\ref{5.5}) by $R_{i,k\ell}$
and integrating over velocity to get the coupled set of
equations
\beq
\label{c21}\left(
\begin{array}{cc}
\tau_{11}+\frac{2\omega^*\xi^{*1/3}}{M_1^{\beta}}& \tau_{12}\\
\tau_{21}&\tau_{22}+\frac{2\omega^*\xi^{*1/3}}{M_2^{\beta}}
\end{array}
\right) \cdot \left(
\begin{array}{c}
\eta_{1}^*\\
\eta_{2}^*
\end{array}
\right)=\left(
\begin{array}{c}
\chi_1^{-1}\\
\chi_2^{-1}
\end{array}
\right).
\eeq
The (reduced) collision frequencies $\tau_{ij}$ are defined by
\begin{equation}
\label{c22}
\tau_{ii}=\frac{1}{(d-1)(d+2)}\frac{1}{n_iT_i^2\nu_0}\int d{\bf
v}_1 R_{i,\alpha\beta} {\cal L}_i\left(f_{i,M}R_{i,\alpha\beta}\right),
\end{equation}
\begin{equation}
\label{c23}
\tau_{ij}=\frac{1}{(d-1)(d+2)}\frac{1}{n_iT_i^2\nu_0}\int d{\bf
v}_1 R_{i,\alpha\beta} {\cal M}_i\left(f_{j,M}R_{j,\alpha\beta}\right),
\end{equation}
where it is understood that $i\neq j$. The evaluation of these collision integrals has been carried out elsewhere \cite{GM07}. Their explicit forms are given by
\begin{eqnarray}
\label{c24}
\tau_{11}&=&\frac{2\pi^{(d-1)/2}}{d(d+2)\Gamma\left(\frac{d}{2}\right)}\left\{
x_1\left(\frac{\sigma_{1}}{\sigma_{12}}\right)^{d-1}(2\theta_1)^{-1/2}
\right.\nonumber\\
& &\times (3+2d-3\alpha_{11})(1+\alpha_{11})+2x_2 \mu_{21}(1+\alpha_{12}) \nonumber\\
& & \times \theta_1^{3/2}\theta_2^{-1/2}\left[
(d+3)(\mu_{12}\theta_2-\mu_{21}\theta_1)\theta_1^{-2}(\theta_1+\theta_2)^{-1/2}\right.\nonumber\\
& &
+\frac{3+2d-3\alpha_{12}}{2}\mu_{21}\theta_1^{-2}(\theta_1+\theta_2)^{1/2}\nonumber\\
& & \left.\left.
+\frac{2d(d+1)-4}{2(d-1)}\theta_1^{-1}(\theta_1+\theta_2)^{-1/2}\right]\right\},
\end{eqnarray}
\begin{eqnarray}
\label{c25}
\tau_{12}&=&\frac{4\pi^{(d-1)/2}}{d(d+2)\Gamma\left(\frac{d}{2}\right)}
x_2\frac{\mu_{21}^2}{\mu_{12}}\theta_1^{3/2}\theta_2^{-1/2}
(1+\alpha_{12})\nonumber\\
& \times&\left[
(d+3)(\mu_{12}\theta_2-\mu_{21}\theta_1)\theta_2^{-2}(\theta_1+\theta_2)^{-1/2}\right.\nonumber\\
& &
+\frac{3+2d-3\alpha_{12}}{2}\mu_{21}\theta_2^{-2}(\theta_1+\theta_2)^{1/2}\nonumber\\
& & \left.
-\frac{2d(d+1)-4}{2(d-1)}\theta_2^{-1}(\theta_1+\theta_2)^{-1/2}\right].
\end{eqnarray}
The expressions for $\tau_{22}$ and $\tau_{21}$ can be obtained by setting $1 \leftrightarrow 2$. The solution of Eq.\ (\ref{c21}) is elementary and yields Eq.\ (\ref{5.29}).

\section{Local stochastic thermostat}
\label{appD}

In this Appendix we display the expressions of the (reduced) diffusion coefficients $D^*$, $D_p^*$, and $D_T^*$ by using a \emph{local} stochastic thermostat ($\partial_t^{(0)}p=\partial_t^{(0)}T=0$). The expressions of the diffusion coefficients \cite{G09} are
\begin{equation}
\label{d1}
D^*=\nu_D^{-1}\left(\chi_1+x_1\frac{\partial \chi_1}{\partial x_{1}}\right),
\end{equation}
\begin{equation}
\label{d2}
D_p^*=\nu_D^{-1}\left(x_1\chi_1-\frac{\rho_1}{\rho}\right), \quad D_T^*=0,
\end{equation}
where $\nu_D$ is given by Eq.\ \eqref{c17}.

\end{document}